\documentclass[journal=jacsat,manuscript=article]{achemso}

\usepackage[version=3]{mhchem} 
\usepackage{amsmath}
\usepackage{textcomp}

\author{Zezhou Liu}
\affiliation{Department of Physics, McGill University, Montréal, Québec}
\author{Wangwei Dong}
\affiliation{Department of Physics, McGill University, Montréal, Québec}
\author{Thomas St-Denis}
\affiliation{Department of Physics, McGill University, Montréal, Québec}
\author{Matheus Azevedo Silva Pessôa}
\affiliation{Department of Physics, McGill University, Montréal, Québec}
\author{Sajad Shiekh}
\affiliation{Department of Physics, McGill University, Montréal, Québec}
\author{Preethi Ravikumar}
\affiliation{Department of Physics, McGill University, Montréal, Québec}
\author{Walter Reisner}
\email{reisner@physics.mcgill.ca}
\affiliation{Department of Physics, McGill University, Montréal, Québec}

\title[An \textsf{achemso} demo]
  {DNA Dynamics in Dual Nanopore Tug-of-War}

\abbreviations{IR,NMR,UV}

\begin{document}







\begin{abstract}
Solid state nanopores have emerged as powerful tools for single-molecule sensing, yet the rapid uncontrolled translocation of the molecule through the pore remains a key limitation.  We have previously demonstrated that an active dual-nanopore system, consisting of two closely spaced pores operated via feedback controlled biasing, shows promise in achieving controlled, slowed-down translocation. Translocation control is achieved via capturing the DNA in a special tug-of-war configuration, whereby opposing electrophoretic forces at each pore are applied to a DNA molecule co-captured at the two pores.  Here, we systematically explore translocation physics during DNA tug-of-war focusing on genomically relevant longer dsDNA using a T$_4$-DNA model (166\,kbp). We find that longer molecules can be trapped in tug-of-war states with an asymmetric partitioning of contour between the pores.  Secondly, we explore the physics of DNA disengagement from a tug-of-war configuration, focusing on the dynamics of DNA free-end escape, in particular how the free-end velocity depends on pore voltage, DNA size and the presence of additional DNA strands between the pores (i.e. arising in the presence of folded translocation).  These findings validate theoretical predictions derived from a first passage model and provide new insight into the physical mechanisms governing molecule disengagement in tug-of-war.

\end{abstract}

\section{Introduction}

In a nanopore sensor, a voltage bias is applied across a nanoscale pore embedded in a thin membrane separating two electrolyte reservoirs. This bias generates a strong electric field that draws analytes through the pore.  As each analyte translocates, it modulates the trans-pore ionic current, creating current fluctuations that encode information relating to specific analyte conformational, structural and chemical properties.   Nanopores have emerged as promising sensors for various biopolymers—including DNA \cite{kasianowicz1996characterization, clarke2009continuous, feng2015nanopore}, proteins \cite{hu2021biological, rosen2014single}, RNA \cite{wanunu2010rapid, workman2019nanopore}, and DNA–protein complexes.\cite{hornblower2007single, squires2015nanopore}   Compared to classic fluorescence based single-molecule approaches, nanopores have the advantage of a purely electrical sensing mechanism that obviates the need for expensive labeling of single molecule species by fluorophores and confers a smaller and more inexpensive device footprint.

A key challenge in achieving a successful nanopore technology is translocation control.\cite{carson2015challenges}  Under a voltage bias sufficient to drive double-stranded DNA (dsDNA) through the pore and generate current signals with an adequate signal-to-noise ratio, the dsDNA chain often translocates at least two orders of magnitude too fast to resolve the single-basepair features.\cite{carson2015challenges} Protein-based pores address the challenge of translocation control via the use of polymerases or helicases fused to the pore that ratchet the DNA through one nucleotide at a time \cite{hu2021biological}, a technology key to achieving nanopore DNA sequencing.  In the context of solid-state pores, several approaches have been considered to achieve translocation control, including modifying the nanopore geometry,\cite{wanunu2012nanopores} adjusting the electrolyte composition,\cite{fologea2005slowing, smeets2006salt} engineering the nanopore surfaces through gating or surface functionalization\cite{luan2012slowing} and optical tweezer based control.\cite{keyser2006direct, leitao2023spatially}

The dual nanopore system is a particularly attractive approach to achieving translocation control in solid-state nanopores.  In a dual nanopore system, a DNA molecule is co-captured in two closely positioned nanopores ($<1$\,$\mu$m spacing).  Reliable co-capture can be facilitated by the use of a field-programmable gate array (FPGA), incorporating active feedback, that dynamically adjusts pore biasing in response to real-time current signals.\cite{CtrlDNATOWSmall2}  Once co-captured, opposing bias voltages can be applied at the two pores, creating a DNA tug-of-war (TOW).  In TOW control, the portion of molecule suspended between the pores is extended and pore-to-pore migration speed is controlled by the differential pore biasing (whereas the current at pore 1 (P1) and pore 2 (P2) is controlled by the absolute voltage bias at each pore).  We demonstrated that dual-nanopore systems can reduce DNA translocation velocity by two orders of magnitude compared to conventional single-nanopore methods.\cite{CtrlDNATOWSmall2} By introducing genomic labeling techniques that attach small protein tags to specific DNA positions, we can perform repeated bi-directional scanning and probe the same DNA region--up to hundreds of times \cite{liu2020flossing, rand2022electronic}.

The formation and maintenance of the TOW state is crucial to achieving enhanced molecular control in the dual-pore system.  Consequently, understanding the physical mechanisms that underlie TOW is crucial for further technology optimization.  Dual-pore tug-of-war has three distinct stages.  The first stage is state formation where the DNA is manipulated by a sequence of feed-back controlled voltage changes to achieve co-capture.  The second stage is state maintenance where the DNA chain is pulled taunt by opposed electrical forces acting at each pore; the electrical forces are sufficiently strong at this stage to fully stretch the DNA strand between the pores.  The third stage is the process of molecule disengagement from TOW.  We previously argued that DNA dynamics during the maintenance stage arises under the joint influence of diffusion and electrophoretic transport and can be quantified by a 1D first-passage theory.\cite{CtrlDNATOWSmall2}   This theory posits a diffusion-convection PDE to describe the exchange of contour between the microchannel reservoirs adjoining P1 and P2.  The model quantifies the time-scale for one of the reservoirs to fully empty of contour leading to molecule disengagement from the adjacent pore (and initiation of stage three).  A key question for the dual-pore technology concerns the physics of the TOW configuration for longer DNA that would facilitate genomic mapping applications ($>100$\,kbp).  Does the first passage model correctly describe TOW with longer DNA?  A second question concerns the physics of the exit stage, initiated when the DNA free end is suddenly released from one pore.  The DNA strand extended between the pores then recoils while the free end transits to and exits from the second pore.  This process is fundamentally of interest as an extreme case of single pore translocation where only the last $\sim 1\%$ of the DNA contour remains on the \textit{cis}-side (e.g. contour present in strand between the pores). With the dual nanopore system, we can measure the transit time of the DNA free end between the pores (i.e., the time-of-flight, TOF) with 4\,$\mu$s precision.  This enables precise measurement of DNA free-end translocation velocity for the last remaining \textit{cis}-side contour. 

To address these questions, we extend the dsDNA TOW experiments to probe T$_4$-DNA, which has a contour length approximately three times that of $\lambda$-DNA used previously.\cite{CtrlDNATOWSmall2}  In these experiments we systematically vary the control voltage to investigate the TOW dwell time distribution response and quantify the DNA free end velocity at the end of the TOW.  We find that the longer T$_4$-DNA compared to $\lambda$-DNA forms asymmetric TOW configurations with unequal partitioning of contour between the reservoirs adjoining P1 and P2.  From analysis of the TOW distributions, we show that the DNA transit velocity between the pores for T$_4$ and $\lambda$-DNA scales linearly with the exiting pore voltage, enabling access to the  electrophoretic mobility for the DNA free end in TOF.   Measurements of free-end dynamics reveal two surprises:  (1) the electrophoretic mobility of the DNA free end is substantially higher than mobilities measured for single and dual pore translocation (by 3 orders of magnitude); (2) despite the similar TOW-end conformations, we observe that the $\lambda$-DNA free end moves roughly twice as fast as the T$_4$-DNA free end, hinting at the possible role of \textit{trans}-side interactions.  Finally, by investigating free end dynamics in transitions from configurations with a single fold to linearized configurations, we explore how the presence of an additional dsDNA filament between the pores alters the free-end dynamics.  We find that the free-end dynamics is slowed in the presence of an additional segment, suggesting that hydrodynamic interactions are present between the adjacent strands.

\section{Results and discussion}

\subsection{Experimental Setup}
\begin{figure}
\includegraphics*[width=\textwidth]{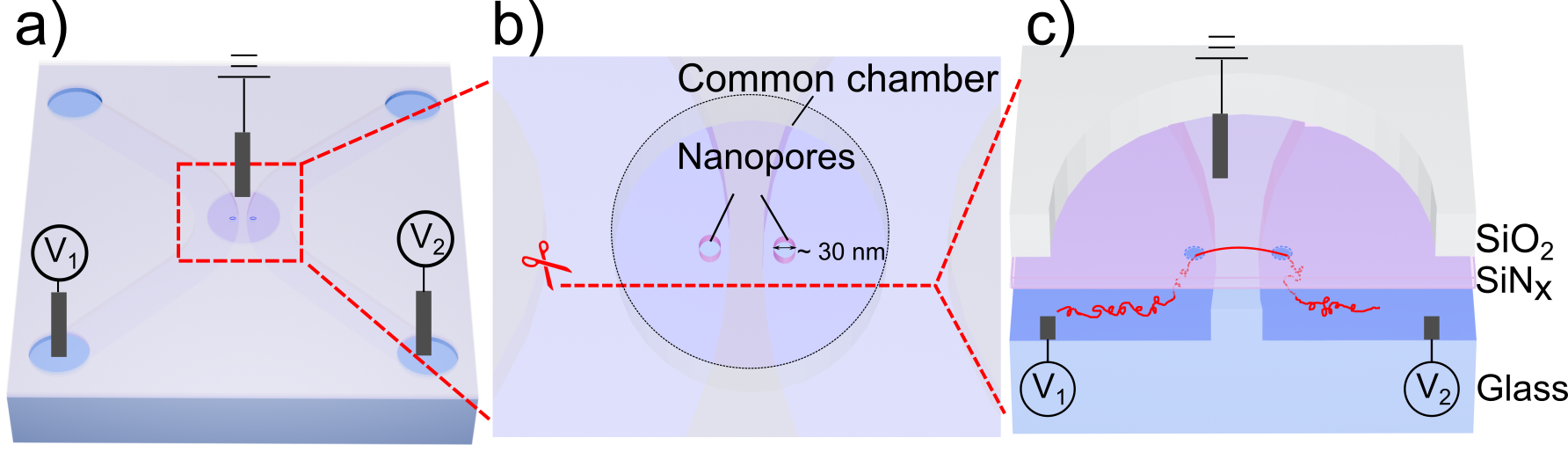}
\caption{\textbf{a)} Schematic of the dual-nanopore device. Voltages V$_1$ and V$_2$ are controlled by the FPGA with the common chamber grounded. \textbf{b)} Zoomed in view and \textbf{c)} cross-section of the region of membrane containing the two nanopores. A cartoon DNA (red) is shown in a tug-of-war (TOW) configuration. The voltages V$_1$ and V$_2$ are set to positive polarity to engage the competing force.} 
\label{fig1:schematic}
\end{figure}

Our dual nanopore device is based on an all-glass chip construction that combines microfluidic channels and nanopore containing membranes to enable independent voltage control and ionic sensing at two closely spaced pores\cite{small2018} (Fig.~\ref{fig1:schematic}). A pair of U-shaped microfluidic channels, embedded in a borosilicate glass substrate, approach each other and reach close proximity in the chip center (see Fig.\,\ref{fig1:schematic}a).  The channels are sealed by a SiN$_x$/SiO$_2$/SiN$_x$ membrane, which is thinned down to $\sim 30$\,nm in a 10\,$\mu$m diameter region at their point of closest approach (see Fig.\,\ref{fig1:schematic}b).  A focused ion beam (FIB) system is then used to drill two nanopores around 30\,nm in diameter, placed $\sim$\,600\,nm apart, in the thinned region (see Fig.\,\ref{fig1:schematic}b).  The space above the two pores, on the membrane facing side of the chip, is interfaced to an isolated external reservoir which we term ``the common chamber''; this is electrically grounded during experiments.  Each pore then connects a given channel to the common chamber.

The channels are electrically biased via Ag/AgCl electrodes inserted in fluidic reservoirs placed at the channel termini.   Voltage is applied and current readings acquired at each pore via a dual-channel patch clamp amplifier interfaced to a field programmable gate array (FPGA).   This setup can perform feedback control of translocation, dynamically adjusting the voltages at each pore ($V_1$, $V_2$) in response to current measurements at P1 and P2. Prior to each experiment, the dsDNA samples used (T$_4$-DNA and $\lambda$-DNA) are diluted to a concentration of $10$\,$\mu$g/ml in nanopore sensing buffer (1\,M LiCl, 10\,mM Tris with 1\,mM EDTA at pH 8.0) and loaded into the common chamber.  We then initiate our active translocation control electronics to capture chains in TOW.  First, the FPGA toggles V$_1$ to positive polarity (+220\,mV) to drive DNA sample through P1. Translocation at P1 is detected when the FPGA senses a 80\,pA absolute current drop relative to baseline.  Following translocation, the FPGA then continuously applies positive V$_1$ for 20\,ms to drive the DNA deeper into the microchannel and further away from P1.  This downwards motion is terminated by zeroing the voltage bias for 10\,ms, allowing the molecule to relax.  DNA re-capture at P1 is initiated by toggling V$_1$ to negative polarity (-220\,mV).   Note that the 20\,ms positive V$_1$ step is required to increase the duration of the re-capture step; this prevents the DNA molecule from translocating through P1 during the time window required to stabilize the capacitive transient arising from reversing the voltage polarity. Following detection of the re-capture translocation at P1, the FPGA maintains V$_1$ negative for 150\,$\mu$s and then sets V$_1$ to zero; this makes available a free DNA ``tail'' in the common chamber sufficiently long for capture by P2 with a positive V$_2$ (+400\,mV).  Following capture of this tail by P2, signaled by detection of a translocation event at P2, the FPGA engages the TOW by setting V$_1$ and V$_2$ to positive values so that opposite forces are applied simultaneously at P1 and P2.  This creates forces in opposing directions that extend the molecule between the pores (see Fig.\,\ref{fig1:schematic}c).

\subsection{DNA Tug-of-War}
\begin{figure}[h]
  \includegraphics*[width=\textwidth]{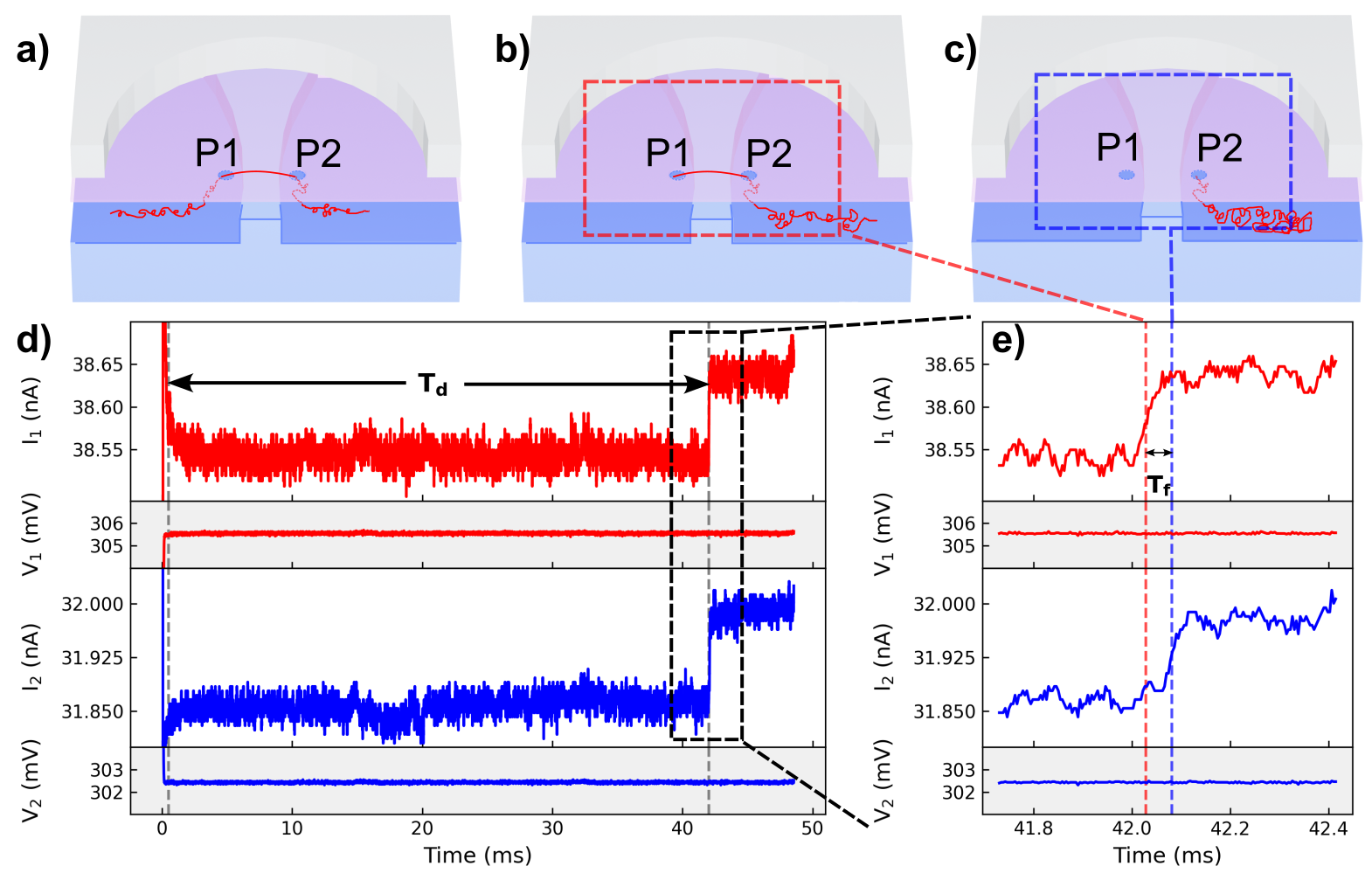}
  \caption{\textbf{a)} The DNA chain is co-captured by the two closely placed nanopores, leading to a TOW event. At the end of the TOW event, \textbf{b)} the DNA free end escapes P1 and then \textbf{c)} escapes P2. \textbf{d)} An example of the DNA TOW events. The grey dashed lines indicate the start and the end of the TOW event; T$_{\mathrm{d}}$ indicates the dwell time of the TOW event. Note that immediately following application of the opposing voltages (i.e. at $t=0$), the capacitive transient leads to an ionic current spike; this stabilizes within 500\,$\mu$s.  The black dashed box indicates the end of the TOW event where the free end escapes from P1 and P2; \textbf{e)} shows a zoom-in view of this portion of the event.  The quantity T$_{\mathrm{f}}$ indicates the time-of-flight for the DNA free end between the pores.}
  \label{fig2:T4tow}
\end{figure}

The TOW event initiates once the DNA molecule is co-captured by P1 and P2 with opposing electrophoretic forces applied at the pores.  The chain in TOW configuration is free to slide between the pores.  This motion is driven both by thermal fluctuations and electrophoretically if the applied forces at the pores are unbalanced (see Fig.\,\ref{fig2:T4tow}a).  The TOW terminates when a DNA free end escapes the left pore (or the right pore) (Fig.\,\ref{fig2:T4tow}b), travels between the pores, and finally translocates through the right pore (or the left pore) (Fig.\,\ref{fig2:T4tow}c).  The ionic current readings from P1 and P2 during the TOW process (Fig.\,\ref{fig2:T4tow}d) can be used to deduce the duration of the TOW configuration (T$_\mathrm{d}$, TOW ``dwell time'').  The TOW configuration is defined as starting 500\,$\mu$s after the voltage is applied; this accounts for the capacitive transient arising upon reversing V$_1$ to initiate TOW.  The TOW state terminates when the chain disengages from one of the two nanopores.  The molecule exit time from a given pore is determined as the time at which the current signal for that pore drops below a threshold set as the average of the corresponding open pore level and level in the presence of translocating DNA (i.e. blockade level).  The time interval between the exit times measured at P1 and P2 (T$_\mathrm{f}$) corresponds to the time-of-flight of the DNA free end (Fig.\,\ref{fig2:T4tow}e).  For T$_4$-DNA, T$_\mathrm{f}$ is on the order of 10\,$\mu$s, approximately 3 orders of magnitude shorter than T$_\mathrm{d}$ ($\sim 10$\,ms).

\subsubsection{Quantifying DNA Dwell Time in Tug-of-War}
\begin{figure}
  \includegraphics*[width=0.85\textwidth]{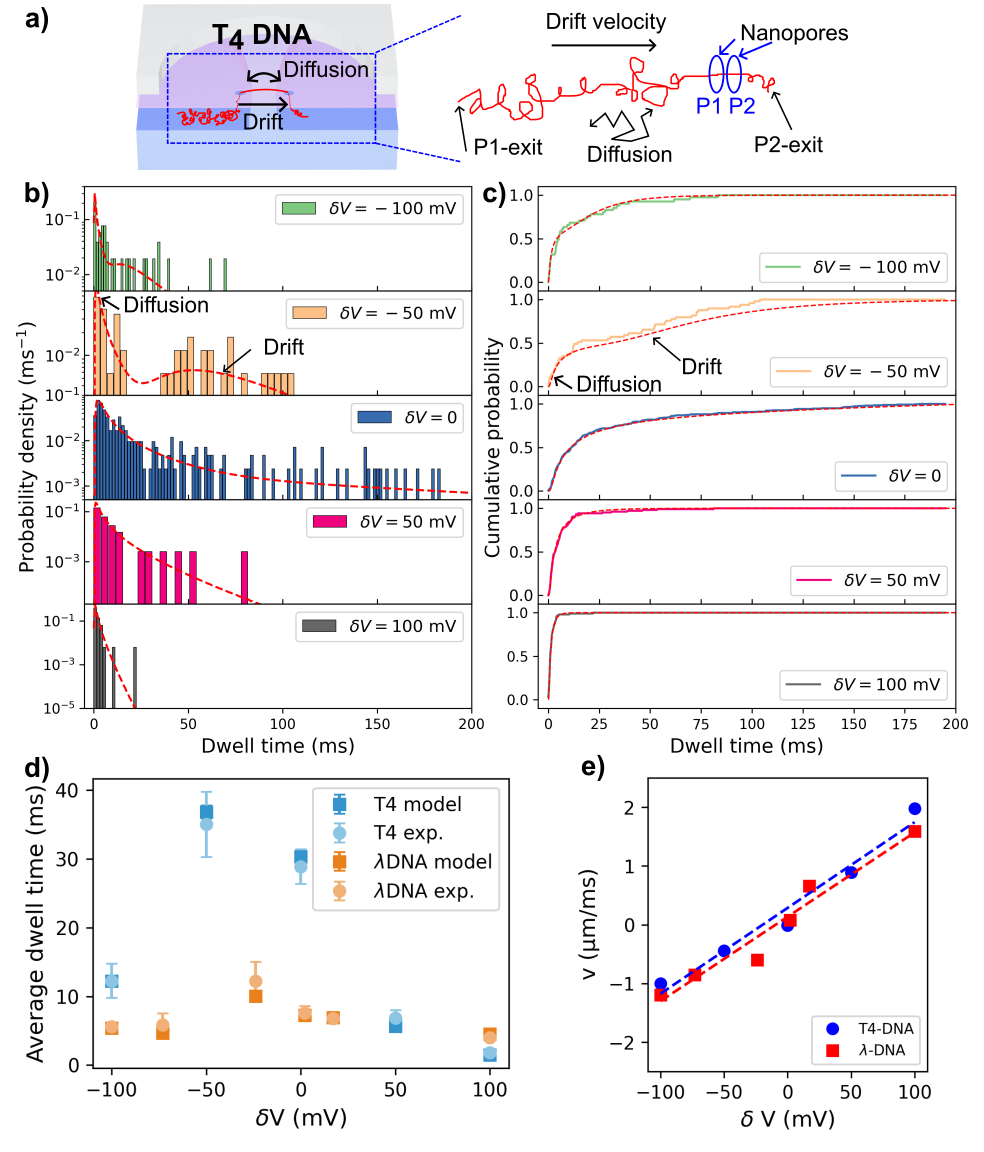}
  \caption{\textbf{a)}. Schematic showing asymmetric initial configuration of the T$_4$-DNA; biasing conditions shown correspond to $V_1<V_2$ (i.e. $\delta V<0$) so the drift velocity points from P1 to P2.  The zoomed-in schematic indicates the polymer free ends that exit at P1 and P2 (we call these free-ends P1-exit and P2-exit).  \textbf{b)}. Histogrammed TOW dwell times for T$_4$-DNA.  The red dashed line indicates fits to the convective diffusion model. The number of events are 84, 48, 260, 104, 98 (for $\delta V$ ranging from -100\,mV to +100\,mV).  \textbf{c)}. The cumulative histogram of TOW dwell-times corresponding to the histograms shown in b). The red dashed line indicates the fits to the convective diffusion model. \textbf{d)} The mean TOW dwell time versus pore voltage differential. The round markers correspond to the experiment and the square markers correspond to the model. The error bars for the experimental average dwell time give the standard error of the mean. The error bars for the model values correspond to the fitting covariance; this is smaller than the marker size. \textbf{e)}. The drift velocity extracted from the convective diffusion model (blue points, T$_4$-DNA; red points $\lambda$-DNA) with accompanying linear fits (dashed lines).}
  \label{fig3:t4dwell_fit}
\end{figure}

The DNA motion in TOW is driven by thermal fluctuations and electrophoretic drift (as depicted in Fig.\,\ref{fig3:t4dwell_fit}a).  We quantify the physics of the TOW state by obtaining the histogram of dwell times for the T$_4$-DNA (see Fig.\,\ref{fig3:t4dwell_fit}b).   As we expect that the electrophoretic force driving sliding of the DNA between the pores should depend on the voltage difference alone, we choose to group the dwell times results based on the voltage difference between the pores $\delta V=V_1-V_2$ rather than the absolute voltages $V_1$ and $V_2$.  This supposition can be demonstrated directly; the average dwell time appears independent of the absolute voltages (see Fig.\,S1 in Supp. Mat.). 

We find that the T$_4$-DNA dwell time distributions depend asymmetrically on the polarity of $\delta$V (i.e. compare $\delta V = \pm50 \mathrm{mV}$ and $\delta V = \pm100 \mathrm{mV}$ in Fig.\,\ref{fig3:t4dwell_fit}b). This dwell distribution asymmetry can be confirmed by inspecting the corresponding cumulative distributions (Fig.\,\ref{fig3:t4dwell_fit}c).  We argue that this asymmetry is a signature of an unbalanced initial TOW position, i.e. the TOW initiates far from the chain center (see Fig.\,\ref{fig3:t4dwell_fit}a).  For $\delta V<0$, the dwell times are very broadly distributed as for these biasing conditions the DNA exits at P1 when the chain free-end P1-exit reaches P1 (Fig.\,\ref{fig3:t4dwell_fit}a), requiring a large amount of contour to thread through P1.  For $\delta V>0$ biasing conditions, the chain exits at P2 when the chain free-end P2-exit reaches P2 (Fig.\,\ref{fig3:t4dwell_fit}a).  In this case, less contour is required to thread through P2; this gives rise to tighter dwell time distributions.  In contrast, $\lambda$-DNA is expected to have a balanced initial position due to its shorter contour length. We observe symmetric dwell distribution for $\lambda$-DNA samples in current work (Fig.\,S2 in Supp. Mat.) and our previous work\cite{CtrlDNATOWSmall2} (see Fig.\,3c,d and Fig.\,S7  in this reference). 

\begin{figure}
    \centering
    \includegraphics[width=\linewidth]{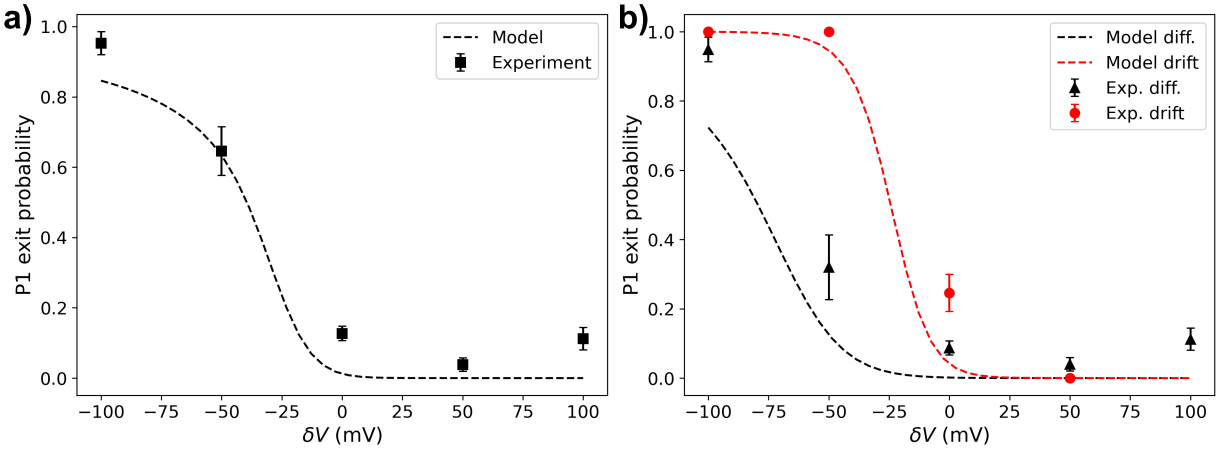}
    \caption{\textbf{a)}\,P1 exit probability for all T$_4$-DNA TOW events as a function of voltage differential. The dashed line is calculated from the first passage model. \textbf{b)} \,P1 exit probability for the T$_4$-DNA TOW events showing the exit probability separately for the diffusion peak (dwell time$<30$\,ms, black triangles) and drift peak (dwell time$\geq30$\,ms, red dots). The error bars give the standard error on the mean.}
    \label{fig4:twopeaks}
\end{figure}

We observe a non-monotonic two-peak structure for the dwell time distribution at $\delta V=-50$\,mV (Fig.\,\ref{fig3:t4dwell_fit}b). The dwell time peaks are located at $\sim 10$\,ms and $\sim 55$\,ms corresponding to two distinct slopes in the cumulative probability.  This two-peak structure is consistent with an unbalanced initial TOW position.  The peak at shorter dwell time corresponds to the DNA exiting at P2 via the free-end P2-exit (Fig.\,\ref{fig3:t4dwell_fit}a); this process is dominated by diffusion.  The peak at longer dwell time corresponds to the DNA exiting at P1 via the free-end P1-exit (Fig.\,\ref{fig3:t4dwell_fit}a); this process is dominated by electrophoretic drift.  The dual-pore setup allows us to directly measure from which pore the chain exits first (i.e. from the first pore current channel to revert to baseline, Fig.~\ref{fig2:T4tow}e). We can then find the P1 exit probability for the two peaks to verify our hypothesis. We group the T$_4$-DNA TOW events by dwell time less/greater than 30\,ms; the resulting P1 exit probability for events in these dwell time bins is shown in Fig.\,\ref{fig4:twopeaks}. We observe for $\delta V=-50$\,mV that the $<30$\,ms bin corresponding to the diffusion peak has a P1 exit probability of around 30\% (i.e. P2 exit probability of 70\%, corresponding to escape from P2); the $>30$\,ms bin corresponding to the drift peak has a P1 exit probability of 100\%.

The dwell time distribution for $\delta V$ close to zero is extended to higher dwell-time values compared to non-zero $\delta V$. We corroborate this observation by determining the mean dwell times for each $\delta V$ (Fig.\,\ref{fig3:t4dwell_fit}d). Note that the mean dwell time is not necessarily maximized exactly at $\delta$V=0. We find that the dwell time is maximized at a negative $\delta$V value that induces a slight drift velocity biasing the chain toward exiting at P1 via free-end P1-exit (Fig.\,\ref{fig3:t4dwell_fit}a).  This slightly counterintuitive effect is also a consequence of how asymmetric partitioning modulates the relative time-scales of diffusional escape (via free-end P2-exit at P2) and escape driven by drift (via free-end P1-exit at P1).  When drift is absent at $\delta$V=0, the dwell time is dictated by diffusional escape of T$_4$-DNA from P2 via free-end P2-exit.  As drift velocity increases in the P1 to P2 direction (i.e. with $\delta$V$<0$), the time-scale of diffusional escape at P2 against the bias direction increases, leading to higher dwell-times.   However, at sufficiently high drift, rapid exit from the free-end P1-exit dominates and the dwell time decreases again.   Our model captures this non-monotonic behavior giving a peak dwell-time at a finite negative drift (see the red dashed line in  Fig.\,S1). In addition, the pores possess slightly different geometries leading to a difference in the detailed distribution of the electric field.\cite{pud2016mechanical}

Following our previous work,\cite{CtrlDNATOWSmall2} we model the dual pore translocation using a 1D convective diffusion equation with two-sided absorbing boundary conditions. In this previous model, we argued that in TOW the stretching forces applied at the pores will maintain a high average tension in the DNA strand linking the pores.  In turn, this high tension leads to a high average fractional extension ($\sim 90$\%) that, at fixed pore separation, fixes the contour present between the pores $x_{cc}$ at roughly a value equal to the interpore spacing (which is less than 5\% of the DNA contour length).  If we let $x_1$ be the DNA contour present in channel 1 and $x_2$ the DNA contour present in channel 2, we expect $x_1$ and $x_2$ to evolve dynamically under the influence of diffusion and a constant electrophoretic force imposed by the voltage difference. The dynamics will terminate when either $x_1$ or $x_2$ reaches zero, signifying emptying of the respective channel of DNA contour. Given the constraint $x_1+x_2+x_{cc}=L$ with $L$ the total DNA contour length, this dynamics can be mapped onto a 1D biased random walk in the contour variable $x \equiv x_1$ with absorbing boundary conditions imposed at $x=0$ and $x=L$.  This problem is analogous to the original biased pore diffusion model proposed by Lubensky and Nelson\cite{lubensky1999driven}, except that in this case absorbing boundary conditions are imposed at both ends of the interval.

The DNA dwell time can be mapped onto the time for $x$ to reach either zero or $L$, formally the first-passage time of 1D biased diffusion with two-sided absorbing boundary conditions.  Formally, we introduce the probability density function (PDF) $P(x,\,t)$ of finding $x$ amount of contour in channel 1.  The PDF satisfies the Smoluchowski equation:
\begin{equation}
  \frac{\partial P(x,t)}{\partial t}=D\frac{\partial^2P(x,t)}{\partial x^2}-v\frac{\partial P(x,t)}{\partial x},
\end{equation}
where $D$ is the diffusion coefficient and $v$ is the drift velocity driven by the voltage bias. The initial condition and boundary conditions correspond to:
\begin{equation}
  \begin{split}
    P(x,0) & = \delta({\alpha}L) \\
    P(0,t) & = P(L,t) = 0.
  \end{split}
\end{equation}
The parameter $\alpha=x_{01}/L$ where $x_{01}$ is the initial starting contour in channel 1.   The distribution of the first-passage time is determined from:
\begin{equation}
  FP(t)=-\frac{d}{dt} \int_{0}^{L}P(s,\,t)ds.
\end{equation}

Note that our previous model\cite{CtrlDNATOWSmall2} included sub-diffusivity.  Sub-diffusivity was necessary in the previous work to model the TOW state at balance point, i.e. when equal and opposite electrophoretic forces were exerted by the pores (leading to $v=0$).  At balance point we found the dwell time distribution had a stretched exponential character with a large number of long-time events approximately one order of magnitude slower than the unbalanced case (see Fig.~3C in our previous work\cite{CtrlDNATOWSmall2}). Here, our experimental data at $\delta V=$0 can still be well fitted by a first-passage model incorporating classic diffusion (see Fig.~\ref{fig2:T4tow}a and Fig.~S2). Regarding why sub-diffusivity does not appear necessary here, one possibility is that sub-diffusivity is reduced in asymmetric partitioning.  Another is that sub-diffusive behavior was linked to the higher degree of stretching present in the previous work (where P2 was set to 500\,mV, compared to 300\,mV used here).

In the first-passage model, we leave the diffusion coefficient $D$, the drift velocity $v$ and the initial fractional contour in channel 1 ($\alpha$) as fitting parameters.  Fitting is performed in python by maximizing an objective function given by the cosine similarity\cite{bishop2006pattern} between the modeled and the experimental dwell time cumulative distribution.  The convective-diffusion equation is solved via the open-source software FreeFEM\cite{MR3043640}. The modeled dwell time cumulative distribution $\mathrm{CDF}_{\mathrm{sim}(t)}$ is evaluated from $1-\frac{\int_{0}^{L}P(x,t)dx}{\int_{0}^{L}P(x,0)dx}$.  We use the Nelder-Mead method to optimize the fitting parameters. Once the model is fitted, we calculate the exit probability by evaluating the model flux $j=-D\frac{dP}{dx}+vP$ at the domain boundary.

Our fitted model recapitulates the asymmetry of the dwell time distribution shown in Fig.\,\ref{fig3:t4dwell_fit}a). From the fitting, we extract the parameter $\alpha$ for the T$_4$-DNA $\alpha_{\mathrm{T4}}=0.92\pm0.02$. This result indicates that $\sim 90\%$ of the T$_4$-DNA-DNA contour is below P1 when TOF initiates. By applying the same TOW setup logic upon the $\lambda$-DNA, the fractional initial position is $\alpha_{\lambda \mathrm{DNA}}=0.44\pm0.08$, which agrees with our previous work\cite{CtrlDNATOWSmall2}. Due to the balanced initial position of the $\lambda$-DNA, the asymmetry of the dwell time distribution is not as significant as the counterpart for the T$_4$-DNA (Fig.\,S2). In addition, our model confirms our hypothesis that the two-peak structure in the dwell-time distribution observed at $\delta V=-50$\,mV is a result of the interplay between the unbalanced initial position and the drift velocity.  Model predictions of the dwell time distribution for asymmetric partitioning and varying drift velocity (Fig.\,S3) show that the two-peak structure appears for increasingly negative $\delta$V biasing when the drift velocity drives the DNA free end P1-exit to escape at P1.   In addition, as the P1 exit probability can be readily obtained from the model by calculating the flux $j$ at the boundary corresponding to P1 escape, exit probability predictions for P1 can be compared to experiment (Fig.\,\ref{fig4:twopeaks}a,b).

We expect that the drift velocity $v$ is proportional to the electrophoretic force difference $f_{\mathrm{ep}}$ between the pores. Given $f_{\mathrm{ep}}=a V$ where $a$ is a parameter that relates the voltage biasing to the force applied at the pore, we have $v=A\delta V$, with A the electrophoretic mobility. The drift velocities are shown in Fig.\,\ref{fig3:t4dwell_fit}e. As expected, the drift velocity depends linearly on $\delta V$ and the electrophoretic mobility for T$_4$-DNA- and $\lambda$-DNA can be extracted via a linear fit. We find $A_{\mathrm{T4-DNA}}=15\pm2\,\mu \mathrm{m}/\mathrm{s}\cdot\mathrm{mV}$ and $A_{\mathrm{\lambda-DNA}}=14\pm2\,\mu\mathrm{m}/\mathrm{s} \cdot \mathrm{mV}$. The results are comparable as expected; $\lambda$-DNA and the T$_4$-DNA should have a very similar charge-to-friction ratio. We can also extract diffusion coefficients for the two cases.  We find $D_{\mathrm{\lambda-DNA}}=4.7\pm0.7\,\mu$m$^2/$ms and $D_{\mathrm{T4-DNA}}=5.6\pm2.4\,\mu$m$^2/$ms. The values of $D$ are approximately four orders of magnitude greater than the bulk diffusion coefficient of T$_4$-DNA in solution ($2.5\times10^{-4}$\,$\mu$m$^2$/ms)\cite{choudiff} and one order of magnitude greater than that observed for $\sim 10$\,nm diameter single solid-state pores via analysis of single pore dwell-time distributions ($\sim$0.15$ \,\mu$m$^2/$ms)\cite{ling2013distribution, carson2014}.  These large values of observed diffusivity could arise from tension propagation dynamics\cite{Sarabadani_2018} that reduce the portion of the chain directly participating in the pore-threading dynamics and consequently the magnitude of the friction factor.  For highly slowed dynamics, such as observed here in the dual-pore setup, the \emph{cis}-side friction factor approaches the limit where only a small portion of the contour very close to the pore participates in the nonequilibrium motion (i.e., corresponding to a tension blob in the pore vicinity).

\subsection{DNA free end time-of-flight between pores}
\begin{figure}[h]
  \includegraphics*[width=\textwidth]{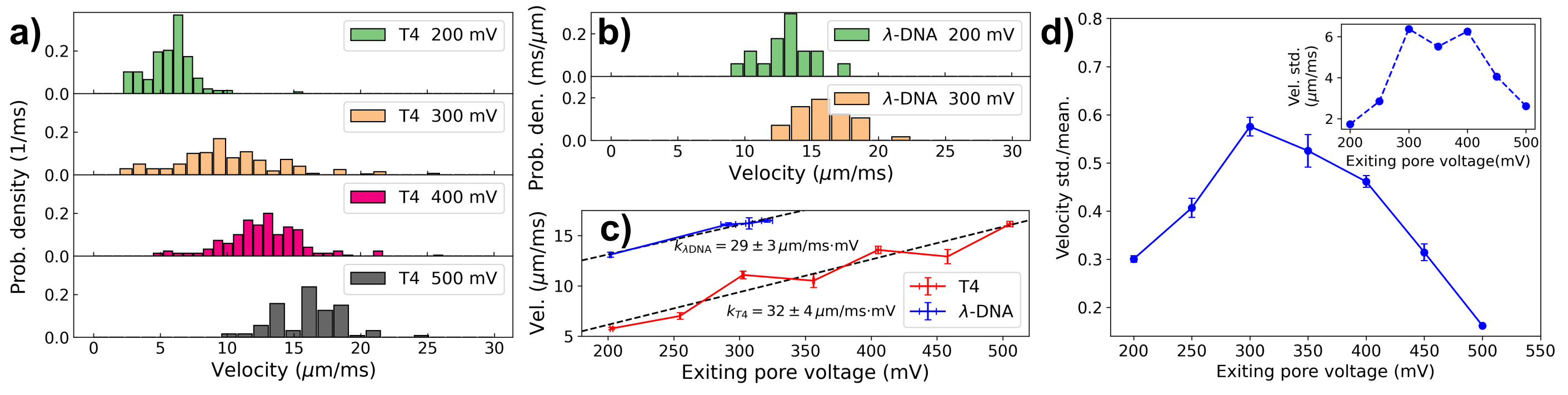}
  \caption{\textbf{a)} Velocity of T$_4$-DNA free end as it travels between the pores as a function of V$_2$. The number of events are 185, 279, 294 and 106 for a P2 voltage of 200\,mV,\,300\,mV,\,400\,mV and 500\,mV. \textbf{b)} Velocity of $\lambda$-DNA free end as it travels between the pores as a function of P2 voltage. The number of events are 51 and 114 for a P2 voltage of 200\,mV and 300\,mV \textbf{c)} Mean velocity of free ends of $\lambda$- and T$_4$-DNA during inter-pore travel as a function of voltage applied to P2. The error bars give standard error on the mean. \textbf{d)} Standard deviation of DNA free end velocity normalized to mean velocity plotted versus P2 voltage. The inset gives the velocity standard deviation value. The error bars are propagated using the standard error on the mean velocity.}
  \label{fig5:end_tof}
\end{figure}

The dual-pore setup enables measurement of the velocity of the DNA free end as it travels between the pores during the disengagement process at the end of the TOW.  We can estimate the free end velocity by dividing the inter-pore distance by T$_\mathrm{f}$ (Fig.\,\ref{fig2:T4tow}e). After the DNA free end escapes P1, the free end's motion is driven by the electrophoretic force applied at P2, proportional to the bias voltage applied at P2. Conversely, for biasing driving motion in the P2 to P1 direction, escape will occur at P2 and the free-end's motion will be determined by the P1 voltage.   We choose to combine these two cases and label the driving voltage in each case ``exiting pore voltage''.  We observe that the velocity of the DNA free end increases as the exiting pore voltage increases (Fig.\,\ref{fig5:end_tof}a,b).  In particular, the mean free end velocity of T$_4$- and $\lambda$-DNA (Fig.\,\ref{fig5:end_tof}c) depends linearly on the exiting pore voltage. The fitted slope $k$ provides a measure of the free end electrophoretic mobility.  We observe that $k_{\lambda-\mathrm{DNA}}=29\pm3$\,$\mu$m/ms$\cdot$mV and $k_{\mathrm{T4-DNA}}=32\pm4$\,$\mu$m/ms$\cdot$mV, indicating the DNA free end for $\lambda$- and T$_4$-DNA have comparable mobility.  Remarkably, the electrophoretic mobility of the DNA free end is $\sim 3$-orders higher than the electrophoretic mobility determined from the TOW configuration. The electrophoretic mobility of a $\lambda$-DNA passing through a single nanopore is approximately $5 \times 10^{4}$\,$\mu$m/ms$\cdot$mV\cite{small2018, kowalczyk2012slowing}, which is only greater than our TOW measurement by a factor of around 2. We expect the DNA in a TOW configuration has a smaller electrophoretic mobility due to the additional friction arising from the second pore (supplemented potentially by hydrodynamic interactions with the nitride surface between the two pores). However, the electrophoretic mobility of the DNA free end is 3 orders of magnitude higher than for both the single pore and the TOW.

One explanation for the increased free-end velocity is reduced \textit{cis}-side friction due to the small amount of \textit{cis}-side contour during free-end translocation. The electrophoretic mobility is proportional to $1/\xi$, where $\xi$ is the chain friction factor, including contributions from the \textit{cis}-side during tension propagation.   We (crudely) estimate a maximum potential \textit{cis}-side Rouse friction during dual-pore and single-pore translocation of $\xi_\mathrm{c}\sim L$, where $L$ is the contour length of the chain.  The maximum friction factor during free end translocation ($\xi_{\mathrm{fe}}$) is proportional to the inter-pore contour, which can be estimated by the distance between the pores (0.6\,$\mu$m).  The ratio $\xi_{\mathrm{c}}/\xi_{\mathrm{fe}} \sim 10^2$, which explains two-orders of magnitude of the observed velocity increase. During free-end translocation, the recoiling of the inter-pore strand and its decreasing length as contour exits P2 will lead to further reductions of $\xi_{\mathrm{fe}}$.  More accurate modeling would need to account for these effects as well as include classic details of \textit{cis}-side tension propagation\cite{Sarabadani_2018} to obtain an accurate estimate of the average value of $\xi_\mathrm{c}$ and make the comparison precise.

We find that, while the free end mobilities of $\lambda$- and T$_4$- DNA are comparable, the $\lambda$-DNA free end velocity values are shifted upwards by about 7.0\,$\mu$m/ms relative to T$_4$-DNA within the voltage range of our experiment (overall, the $\lambda$-DNA free end travels $\sim 2.5 \times$ faster).  We find this velocity difference somewhat surprising; differences due to overall molecule size can only arise from differences in the amount of contour transferred to the \emph{trans} side of the exiting pore, suggesting that the velocity difference might arise from \textit{trans}-side dynamics.  One possibility is that the T$_4$-DNA, due to its larger size and slower relaxation, might tend to have a greater accumulation of DNA near the pore outlet than the case of $\lambda$-DNA.  This might lead to a slow-down of the translocation dynamics at the exiting pore due to the free energy cost of inserting extra DNA into an already dense DNA packing at the pore outlet (i.e. this free energy cost arises from DNA self-exclusion, analogous to the repulsive electrostatic interactions arising in loading nanofluidic cavities by DNA\cite{reisnerpit,klotzpit}).  Another possibility is that differences in the \textit{trans}-side DNA packing might subtly alter \textit{trans}-side frictional forces.

We also observe significant fluctuations of the free end velocity (Fig.~\ref{fig5:end_tof}d). In single pore studies the variance of DNA translocation velocity is often attributed to the fluctuation of the DNA \textit{cis}-side conformation\cite{storm2005fast,lu2011origins,plesa2015velocity}. However, during the time-of-flight the DNA on the \textit{cis}-side, i.e. in the common chamber near the exiting pore, is in a relatively low entropy configuration as the strand is initially fully stretched (i.e. right after the system escapes from tug-of-war).  We suggest that the free end velocity fluctuations arise from two sources:  (1) hydrodynamic drag of the \textit{cis}-side DNA and (2) extensional fluctuations of the relaxing strand in the common chamber that vary overall friction and chain tension at the exiting pore and lead to consequent variation of translocation rate.  We observe in particular that the DNA free end velocity fluctuations normalized to mean velocity ($\sqrt{\langle (\delta v)^2 \rangle}/\langle v\rangle$) rise, peak around 300\,mV and then fall (see Fig.\,\ref{fig5:end_tof}d). The very simplest models describing translocation velocity fluctuations in pores\cite{lu2011origins, carson2014} assume a constant diffusivity; first-passage theory applied to a single pore\cite{ling2013distribution} then yields a normalized velocity fluctuation standard deviation that is inversely proportional to voltage.  While some single nanopore studies are consistent with this simple prediction,\cite{storm2005fast, fologea2005slowing} others show more complex behavior.  D. Ling \textit{et al.}\cite{ling2013distribution} in $\sim10$\,nm pores observe that the diffusivity is not constant, but varies non-monotonically with voltage; the diffusivity decreases till 50\,mV and then begins to rapidly increase with a quadratic scaling on voltage. This behavior is attributed to fluctuations induced as the  translocating molecule samples regions of varying electroosmotic flow in the pore.  S. Carson \emph{et al}\cite{carson2014} observe a slight increase of diffusivity with voltage for translocation of 500\,bp DNA through 3\,nm pores in the 200-350\,mV range.  Deducing the origins of velocity fluctuations in our system during free-end transit to pinpoint the origin of the nonmonotonic behavior will likely require detailed simulations to couple pore threading with the relaxation dynamics as the free-end recoils towards P2.  

\subsubsection{DNA free end velocity in presence of folding}
\begin{figure}
  \includegraphics*[width=\textwidth]{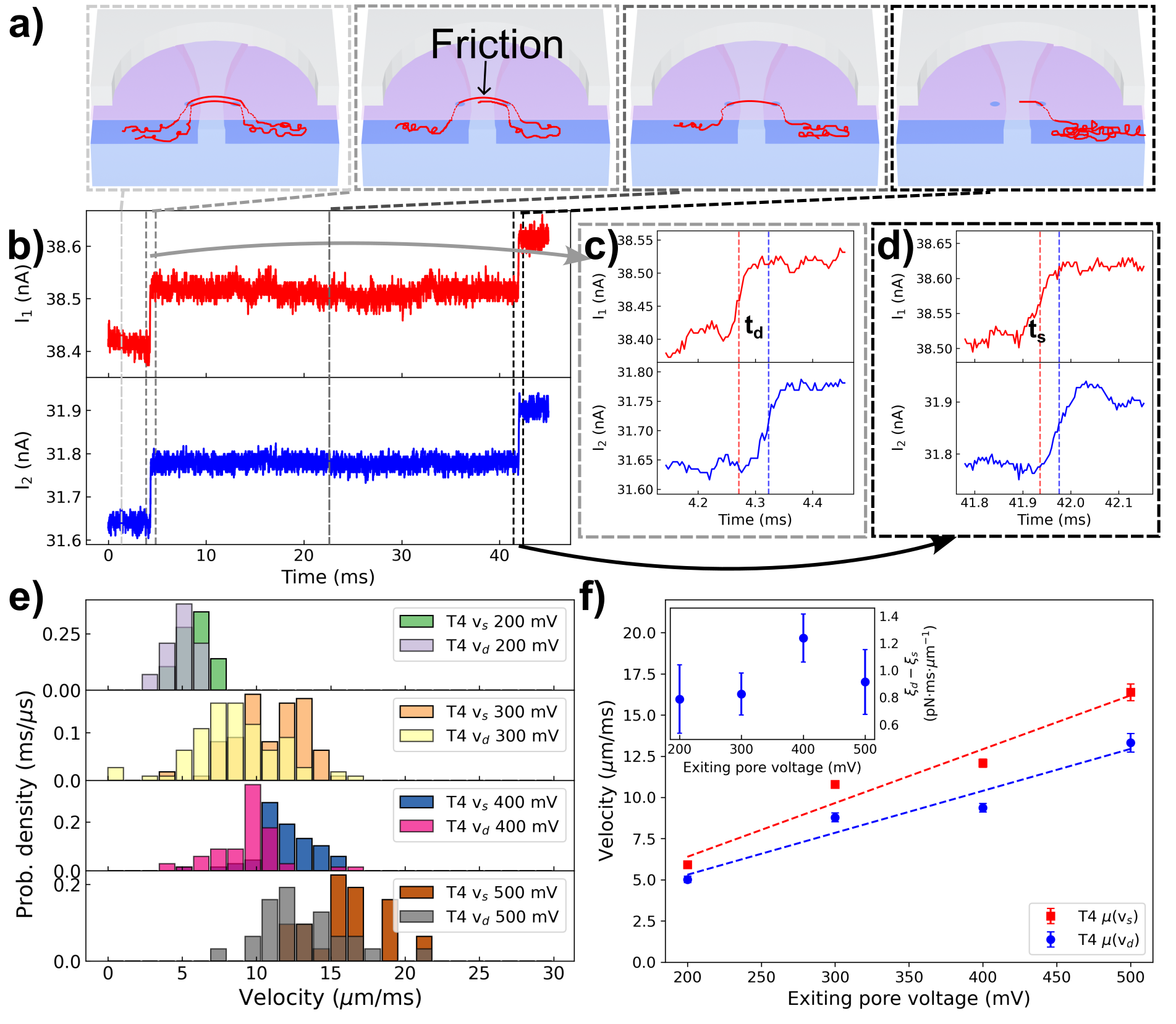}
  \caption{\textbf{a)} Schematic of a folded DNA exiting two pores sequentially, with removal of a fold leading to a linearized conformation that then disengages.  \textbf{b)} Current trace for P1 and P2 from a folded T$_4$-DNA chain. \textbf{c)}. The double filament DNA free end time delay t$_d$ and \textbf{d)} single filament DNA free end time delay t$_s$. \textbf{e)}. The average velocity of the T$_4$-DNA free end passing from P$_1$ to P$_2$, with or without the presence of the second DNA filament. The number of events are 25, 96, 59, 27 for exiting pore voltages ranging from 200\,mV to 500\,mV. \textbf{f)} The velocity of the DNA free end with ($v_{\mathrm{d}}$) and without ($v_{\mathrm{s}}$) a second DNA strand. The dashed line shows linear fits used to extract estimated friction factors. The inset gives the friction coefficient difference $\xi_d-\xi_s$.  The error bars give standard error of the mean.}
  \label{fig6:fold}
\end{figure}

DNA chains can be folded during the TOW process. Folding can be detected via the presence of multiple fixed ratio steps in the ionic current during DNA translocation (Fig.\,\ref{fig6:fold}a).  We consider only configurations with single fold as we detect relatively few events with multiple folding (less than 40 events per folding configuration, see Fig.\,S5).  For a TOW event possessing a single fold, the DNA can unfold by passing the DNA free end from P1 to P2.  The unfolding process can be monitored from the current traces by observing when the current level increases from a two-level blockade to a single-level blockade (Fig.\,\ref{fig6:fold}b).  By separately observing the transitions between a two-level to one-level blockade, and from one-level to open pore, we can obtain the velocity of the DNA free end alone and in the presence of an adjacent DNA strand from a folded portion of the same molecule.  Figure~\ref{fig6:fold}c shows the DNA free end velocity distributions for the folded ($v_{\mathrm{d}}$) and unfolded ($v_{\mathrm{s}}$) configurations. We find that DNA free end consistently transits faster in the unfolded configuration than in the folded configuration (Fig.\,\ref{fig6:fold}e, Fig.\,S4). We attribute the lower velocity of the free end in the folded configuration to the additional friction introduced by the adjacent dsDNA filament in the common chamber and inside the nanopore. 

We calculate the electrophoretic mobility of the DNA free end via linear fitting to the velocity versus voltage results in Fig.\,\ref{fig6:fold}f. The electrophoretic mobility for the unfolded configuration $k_{\mathrm{T4-DNA},{\mathrm{s}}}=33\pm5$\,$\mu$m/ms$\cdot$mV and for the folded configuration $k_{\mathrm{T4-DNA},{\mathrm{d}}}=25\pm5$\,$\mu$m/ms$\cdot$mV.  We attribute the electrophoretic mobility difference between the folded and unfolded cases to the increase of the hydrodynamic friction introduced by the presence of a second DNA strand from the fold. We estimate the hydrodynamic friction coefficient $\xi$ by assuming a force balance between the electrophoretic driving force $f_{\mathrm{ep}}$ and the hydrodynamic friction $f_{\mathrm{r}}$, leading to $f_{\mathrm{ep}}=f_{\mathrm{r}}=\xi v$ where $v$ is the velocity of the DNA free end.  Measurements of nanopore induced electrophoretic force via optical tweezers suggest that pores of the size used here have a force per voltage calibration of 0.13\,pN mV$^{-1}$.\cite{keyser2006direct}.  This suggests the friction coefficient for the DNA free end in the unfolded configuration $\xi_{s}=4.1\pm0.3$\,pN$\cdot $ms $\cdot \mu$m$^{-1}$ and in the folded $\xi_{d}=5.0\pm 0.4$\,pN$\cdot$ms $\cdot \mu$m$^{-1}$.  The difference in friction between the folded and unfolded case $\xi_{\mathrm{d}}-\xi_{\mathrm{s}}$ appears independent of voltage over the range 200-500\,mV (Fig.\,\ref{fig6:fold}f, inset).

\section{Discussion and Conclusion}

We capture DNA molecules in a dual-pore TOW configuration and explore the physics of TOW, in particular focusing on how using larger DNA alters the TOW dynamics.  Comparing T$_4$-DNA to $\lambda$-DNA, we observe that the larger T$_4$-DNA (166\,kbp versus 48.5\,kbp) gives rise to dwell time distributions that show substantial asymmetry as a function of voltage.   We argue that this asymmetry arises from an unbalanced initial position in the TOW configuration. This unbalanced initial position also yields more complicated structures in the dwell-time distribution, including distinct peaks relating to diffusional and drift related escape processes.  This finding is relevant for optimizing the dual-nanopore technology to analyze genomically relevant DNA samples that include DNA fragments of variable length.  For molecules of variable length, the TOW may not always initiate at the molecule center for given control conditions.  Our ability to measure the molecule electrophoretic mobility during TOW suggests that the dual-pore system could potentially distinguish biopolymer samples based on mobility differences, complementing the classical approach of comparing their nanopore current traces. In the future, we can explore DNA samples with different charge properties, for example DNA coated with protein diblock copolymers,\cite{hernandez2012coating} to test the sensitivity of the electrophoretic mobility measurement with TOW. 

In addition we measure the DNA free end velocity during molecule disengagement from the TOW.   The translocation of the DNA free end is an extreme case of single pore translocation where the small amount of DNA contour on the \textit{cis}-side (arising from the linker connecting P1 and P2) rapidly translates through P2.  We find that the DNA free end electrophoretic mobility is approximately 3-orders of magnitude higher compared with TOW and single pore translocation. We attribute this difference to the reduced \textit{cis}-side DNA configuration in free-end translocation that leads to lowered friction.  In addition,  we observe a faster DNA free end translocation for the $\lambda$-DNA compared with the T$_4$-DNA.  We suspect that the larger T$_4$-DNA may accumulate at the pore on the \textit{trans}-side compared to the shorter $\lambda$-DNA, leading to crowding effects that could slow-down translation of the larger molecule. In the future, we can vary the \textit{trans}-side DNA configuration by varying the buffer ionic strength to vary DNA self-interaction strength (i.e. effective width\cite{klotzpit}) and adding additional crowding agent. 

The DNA free end velocity \textit{always} decreases when a second dsDNA strand is present inside the pore (see Fig.\,S4), indicating additional friction arising from the folded structure.  N. Laohakunakorn \textit{et al.}\cite{laohakunakorn2013dna} measured the friction of multiple dsDNAs inside a nanopore by tethering the DNAs to a bead held in place by optical tweezes.  They demonstrated that the hydrodynamic drag increases as the number of DNA molecules increases (see Fig.\,S2).  Note that our experiment is intrinsically different as it includes the relative motion (sliding) between the two dsDNA strands, i.e. the free end of the DNA and the opposing strand within the fold. In N. Laohakunakorn’s work,\cite{laohakunakorn2013dna} this relative motion was neglected because DNAs are tethered to the same bead. In our experiment, we observed that the velocity of the DNA free end is about 10\,$\mu$m/ms, whereas the drift velocity of DNA in the TOW configuration is below 1\,$\mu$m/ms. This indicates that the DNA free end effectively slides against the other strand, which remains approximately fixed. In the future, we can study how the degree of relative motion between the DNA strands affects the DNA free end friction by assessing free end dynamics in more complicated folds.

\section{Experimental methods}
\subsection{Fabrication of the dual-nanopore device}
The detailed fabrication process of the nanopore chips is described in Ref.~\cite{small2018}. Two U-shaped microchannels are dry-etched into a glass wafer, with their tips $0.4$\,$\mu$m apart. A multilayer stack is deposited on a Si wafer: 400\,nm SiN using low-pressure chemical vapor deposition (LPCVD), 100\,nm \ce{SiO2} using plasma-enhanced chemical vapor deposition (PECVD), and 30\,nm SiN using LPCVD. After anodic bonding to the glass wafer, the backside SiN is dry-etched, and the Si substrate  removed using hot KOH. A 10\,$\mu$m$\times10$\,$\mu$m window is opened by dry-etching through the top SiN and partially into the oxide, followed by wet etching of the remaining \ce{SiO2} in hot KOH. Finally, two nanopores are drilled in the SiN membrane at the channel tips using focused ion beam (FIB). 
 
\subsection{DNA samples}
The $\lambda$-DNA (cat. N3011S) is commercially available from the New England Biolabs\textregistered and the T4 GT7 DNA (cat. 318-03971) is available from FUJIFILM Wako Chemicals U.S.A. Corporation. The DNA samples are diluted to approximately 50\,pM in nanopore sensing buffer (1M LiCl 1$\times$TE at pH=8.0) right before the experiment.

\subsection{Nanopore measurement}
The channels and common chamber were first filled with nanopore sensing buffer using a pipette and capillary force. Following testing the baselines of the two pores for noise quality, diluted DNA samples are added to the common chamber using a pipette. A dual-channel voltage- clamp amplifier (MultiClamp 700B, Molecular Devices, Sunnyvale, CA) is used to apply transmembrane voltage and measure ionic current, with the four-pole Bessel filter set at 10\,kHz. A digitizer (Digidata 1440A, Molecular Devices) stored data sampled at 250\,kHz without filtering. The active control is achieved by using logic as explained in the main text, programmed in Labview (2017) and executed on an FPGA (National Instruments model PCIe-7851).

\subsection{Acknowledgments}
We acknowledge support from the Natural Science and Engineering Research Council of Canada (NSERC RGPIN-2018-06125), the Fonds de recherche du Quebec-Nature et technologies (FRQNT-NT, PR-286442) and National Institute of Health (NIH) award number 1R21HG011236-01.

\bibliography{reference}

\providecommand{\latin}[1]{#1}
\makeatletter
\providecommand{\doi}
  {\begingroup\let\do\@makeother\dospecials
  \catcode`\{=1 \catcode`\}=2 \doi@aux}
\providecommand{\doi@aux}[1]{\endgroup\texttt{#1}}
\makeatother
\providecommand*\mcitethebibliography{\thebibliography}
\csname @ifundefined\endcsname{endmcitethebibliography}
  {\let\endmcitethebibliography\endthebibliography}{}
\begin{mcitethebibliography}{37}
\providecommand*\natexlab[1]{#1}
\providecommand*\mciteSetBstSublistMode[1]{}
\providecommand*\mciteSetBstMaxWidthForm[2]{}
\providecommand*\mciteBstWouldAddEndPuncttrue
  {\def\EndOfBibitem{\unskip.}}
\providecommand*\mciteBstWouldAddEndPunctfalse
  {\let\EndOfBibitem\relax}
\providecommand*\mciteSetBstMidEndSepPunct[3]{}
\providecommand*\mciteSetBstSublistLabelBeginEnd[3]{}
\providecommand*\EndOfBibitem{}
\mciteSetBstSublistMode{f}
\mciteSetBstMaxWidthForm{subitem}{(\alph{mcitesubitemcount})}
\mciteSetBstSublistLabelBeginEnd
  {\mcitemaxwidthsubitemform\space}
  {\relax}
  {\relax}

\bibitem[Kasianowicz \latin{et~al.}(1996)Kasianowicz, Brandin, Branton, and
  Deamer]{kasianowicz1996characterization}
Kasianowicz,~J.~J.; Brandin,~E.; Branton,~D.; Deamer,~D.~W. Characterization of
  individual polynucleotide molecules using a membrane channel.
  \emph{Proceedings of the National Academy of Sciences} \textbf{1996},
  \emph{93}, 13770--13773\relax
\mciteBstWouldAddEndPuncttrue
\mciteSetBstMidEndSepPunct{\mcitedefaultmidpunct}
{\mcitedefaultendpunct}{\mcitedefaultseppunct}\relax
\EndOfBibitem
\bibitem[Clarke \latin{et~al.}(2009)Clarke, Wu, Jayasinghe, Patel, Reid, and
  Bayley]{clarke2009continuous}
Clarke,~J.; Wu,~H.-C.; Jayasinghe,~L.; Patel,~A.; Reid,~S.; Bayley,~H.
  Continuous base identification for single-molecule nanopore DNA sequencing.
  \emph{Nature nanotechnology} \textbf{2009}, \emph{4}, 265--270\relax
\mciteBstWouldAddEndPuncttrue
\mciteSetBstMidEndSepPunct{\mcitedefaultmidpunct}
{\mcitedefaultendpunct}{\mcitedefaultseppunct}\relax
\EndOfBibitem
\bibitem[Feng \latin{et~al.}(2015)Feng, Zhang, Ying, Wang, and
  Du]{feng2015nanopore}
Feng,~Y.; Zhang,~Y.; Ying,~C.; Wang,~D.; Du,~C. Nanopore-based
  fourth-generation DNA sequencing technology. \emph{Genomics, proteomics \&
  bioinformatics} \textbf{2015}, \emph{13}, 4--16\relax
\mciteBstWouldAddEndPuncttrue
\mciteSetBstMidEndSepPunct{\mcitedefaultmidpunct}
{\mcitedefaultendpunct}{\mcitedefaultseppunct}\relax
\EndOfBibitem
\bibitem[Hu \latin{et~al.}(2021)Hu, Huo, Ying, and Long]{hu2021biological}
Hu,~Z.-L.; Huo,~M.-Z.; Ying,~Y.-L.; Long,~Y.-T. Biological nanopore approach
  for single-molecule protein sequencing. \emph{Angewandte Chemie}
  \textbf{2021}, \emph{133}, 14862--14873\relax
\mciteBstWouldAddEndPuncttrue
\mciteSetBstMidEndSepPunct{\mcitedefaultmidpunct}
{\mcitedefaultendpunct}{\mcitedefaultseppunct}\relax
\EndOfBibitem
\bibitem[Rosen \latin{et~al.}(2014)Rosen, Rodriguez-Larrea, and
  Bayley]{rosen2014single}
Rosen,~C.~B.; Rodriguez-Larrea,~D.; Bayley,~H. Single-molecule site-specific
  detection of protein phosphorylation with a nanopore. \emph{Nature
  biotechnology} \textbf{2014}, \emph{32}, 179--181\relax
\mciteBstWouldAddEndPuncttrue
\mciteSetBstMidEndSepPunct{\mcitedefaultmidpunct}
{\mcitedefaultendpunct}{\mcitedefaultseppunct}\relax
\EndOfBibitem
\bibitem[Wanunu \latin{et~al.}(2010)Wanunu, Dadosh, Ray, Jin, McReynolds, and
  Drndi{\'c}]{wanunu2010rapid}
Wanunu,~M.; Dadosh,~T.; Ray,~V.; Jin,~J.; McReynolds,~L.; Drndi{\'c},~M. Rapid
  electronic detection of probe-specific microRNAs using thin nanopore sensors.
  \emph{Nature nanotechnology} \textbf{2010}, \emph{5}, 807--814\relax
\mciteBstWouldAddEndPuncttrue
\mciteSetBstMidEndSepPunct{\mcitedefaultmidpunct}
{\mcitedefaultendpunct}{\mcitedefaultseppunct}\relax
\EndOfBibitem
\bibitem[Workman \latin{et~al.}(2019)Workman, Tang, Tang, Jain, Tyson, Razaghi,
  Zuzarte, Gilpatrick, Payne, Quick, \latin{et~al.}
  others]{workman2019nanopore}
Workman,~R.~E.; Tang,~A.~D.; Tang,~P.~S.; Jain,~M.; Tyson,~J.~R.; Razaghi,~R.;
  Zuzarte,~P.~C.; Gilpatrick,~T.; Payne,~A.; Quick,~J.; others Nanopore native
  RNA sequencing of a human poly (A) transcriptome. \emph{Nature methods}
  \textbf{2019}, \emph{16}, 1297--1305\relax
\mciteBstWouldAddEndPuncttrue
\mciteSetBstMidEndSepPunct{\mcitedefaultmidpunct}
{\mcitedefaultendpunct}{\mcitedefaultseppunct}\relax
\EndOfBibitem
\bibitem[Hornblower \latin{et~al.}(2007)Hornblower, Coombs, Whitaker,
  Kolomeisky, Picone, Meller, and Akeson]{hornblower2007single}
Hornblower,~B.; Coombs,~A.; Whitaker,~R.~D.; Kolomeisky,~A.; Picone,~S.~J.;
  Meller,~A.; Akeson,~M. Single-molecule analysis of DNA-protein complexes
  using nanopores. \emph{Nature methods} \textbf{2007}, \emph{4},
  315--317\relax
\mciteBstWouldAddEndPuncttrue
\mciteSetBstMidEndSepPunct{\mcitedefaultmidpunct}
{\mcitedefaultendpunct}{\mcitedefaultseppunct}\relax
\EndOfBibitem
\bibitem[Squires \latin{et~al.}(2015)Squires, Atas, and
  Meller]{squires2015nanopore}
Squires,~A.; Atas,~E.; Meller,~A. Nanopore sensing of individual transcription
  factors bound to DNA. \emph{Scientific reports} \textbf{2015}, \emph{5},
  11643\relax
\mciteBstWouldAddEndPuncttrue
\mciteSetBstMidEndSepPunct{\mcitedefaultmidpunct}
{\mcitedefaultendpunct}{\mcitedefaultseppunct}\relax
\EndOfBibitem
\bibitem[Carson and Wanunu(2015)Carson, and Wanunu]{carson2015challenges}
Carson,~S.; Wanunu,~M. Challenges in DNA motion control and sequence readout
  using nanopore devices. \emph{Nanotechnology} \textbf{2015}, \emph{26},
  074004\relax
\mciteBstWouldAddEndPuncttrue
\mciteSetBstMidEndSepPunct{\mcitedefaultmidpunct}
{\mcitedefaultendpunct}{\mcitedefaultseppunct}\relax
\EndOfBibitem
\bibitem[Wanunu(2012)]{wanunu2012nanopores}
Wanunu,~M. Nanopores: A journey towards DNA sequencing. \emph{Physics of life
  reviews} \textbf{2012}, \emph{9}, 125--158\relax
\mciteBstWouldAddEndPuncttrue
\mciteSetBstMidEndSepPunct{\mcitedefaultmidpunct}
{\mcitedefaultendpunct}{\mcitedefaultseppunct}\relax
\EndOfBibitem
\bibitem[Fologea \latin{et~al.}(2005)Fologea, Uplinger, Thomas, McNabb, and
  Li]{fologea2005slowing}
Fologea,~D.; Uplinger,~J.; Thomas,~B.; McNabb,~D.~S.; Li,~J. Slowing DNA
  translocation in a solid-state nanopore. \emph{Nano letters} \textbf{2005},
  \emph{5}, 1734--1737\relax
\mciteBstWouldAddEndPuncttrue
\mciteSetBstMidEndSepPunct{\mcitedefaultmidpunct}
{\mcitedefaultendpunct}{\mcitedefaultseppunct}\relax
\EndOfBibitem
\bibitem[Smeets \latin{et~al.}(2006)Smeets, Keyser, Krapf, Wu, Dekker, and
  Dekker]{smeets2006salt}
Smeets,~R.~M.; Keyser,~U.~F.; Krapf,~D.; Wu,~M.-Y.; Dekker,~N.~H.; Dekker,~C.
  Salt dependence of ion transport and DNA translocation through solid-state
  nanopores. \emph{Nano letters} \textbf{2006}, \emph{6}, 89--95\relax
\mciteBstWouldAddEndPuncttrue
\mciteSetBstMidEndSepPunct{\mcitedefaultmidpunct}
{\mcitedefaultendpunct}{\mcitedefaultseppunct}\relax
\EndOfBibitem
\bibitem[Luan \latin{et~al.}(2012)Luan, Stolovitzky, and
  Martyna]{luan2012slowing}
Luan,~B.; Stolovitzky,~G.; Martyna,~G. Slowing and controlling the
  translocation of DNA in a solid-state nanopore. \emph{Nanoscale}
  \textbf{2012}, \emph{4}, 1068--1077\relax
\mciteBstWouldAddEndPuncttrue
\mciteSetBstMidEndSepPunct{\mcitedefaultmidpunct}
{\mcitedefaultendpunct}{\mcitedefaultseppunct}\relax
\EndOfBibitem
\bibitem[Keyser \latin{et~al.}(2006)Keyser, Koeleman, Van~Dorp, Krapf, Smeets,
  Lemay, Dekker, and Dekker]{keyser2006direct}
Keyser,~U.~F.; Koeleman,~B.~N.; Van~Dorp,~S.; Krapf,~D.; Smeets,~R.~M.;
  Lemay,~S.~G.; Dekker,~N.~H.; Dekker,~C. Direct force measurements on DNA in a
  solid-state nanopore. \emph{Nature Physics} \textbf{2006}, \emph{2},
  473--477\relax
\mciteBstWouldAddEndPuncttrue
\mciteSetBstMidEndSepPunct{\mcitedefaultmidpunct}
{\mcitedefaultendpunct}{\mcitedefaultseppunct}\relax
\EndOfBibitem
\bibitem[Leitao \latin{et~al.}(2023)Leitao, Navikas, Miljkovic, Drake, Marion,
  Pistoletti~Blanchet, Chen, Mayer, Keyser, Kuhn, \latin{et~al.}
  others]{leitao2023spatially}
Leitao,~S.~M.; Navikas,~V.; Miljkovic,~H.; Drake,~B.; Marion,~S.;
  Pistoletti~Blanchet,~G.; Chen,~K.; Mayer,~S.~F.; Keyser,~U.; Kuhn,~A.; others
  Spatially multiplexed single-molecule translocations through a nanopore at
  controlled speeds. \emph{Nature Nanotechnology} \textbf{2023}, \emph{18},
  1078--1084\relax
\mciteBstWouldAddEndPuncttrue
\mciteSetBstMidEndSepPunct{\mcitedefaultmidpunct}
{\mcitedefaultendpunct}{\mcitedefaultseppunct}\relax
\EndOfBibitem
\bibitem[Liu \latin{et~al.}(2019)Liu, Zhang, Nagel, Reisner, and
  Dunbar]{CtrlDNATOWSmall2}
Liu,~X.; Zhang,~Y.; Nagel,~R.; Reisner,~W.; Dunbar,~W.~B. Controlling DNA
  Tug-of-War in a Dual Nanopore Device. \emph{Small} \textbf{2019}, \emph{15},
  1901704\relax
\mciteBstWouldAddEndPuncttrue
\mciteSetBstMidEndSepPunct{\mcitedefaultmidpunct}
{\mcitedefaultendpunct}{\mcitedefaultseppunct}\relax
\EndOfBibitem
\bibitem[Liu \latin{et~al.}(2020)Liu, Zimny, Zhang, Rana, Nagel, Reisner, and
  Dunbar]{liu2020flossing}
Liu,~X.; Zimny,~P.; Zhang,~Y.; Rana,~A.; Nagel,~R.; Reisner,~W.; Dunbar,~W.~B.
  Flossing DNA in a dual nanopore device. \emph{Small} \textbf{2020},
  \emph{16}, 1905379\relax
\mciteBstWouldAddEndPuncttrue
\mciteSetBstMidEndSepPunct{\mcitedefaultmidpunct}
{\mcitedefaultendpunct}{\mcitedefaultseppunct}\relax
\EndOfBibitem
\bibitem[Rand \latin{et~al.}(2022)Rand, Zimny, Nagel, Telang, Mollison, Bruns,
  Leff, Reisner, and Dunbar]{rand2022electronic}
Rand,~A.; Zimny,~P.; Nagel,~R.; Telang,~C.; Mollison,~J.; Bruns,~A.; Leff,~E.;
  Reisner,~W.~W.; Dunbar,~W.~B. Electronic mapping of a bacterial genome with
  dual solid-state nanopores and active single-molecule control. \emph{ACS
  nano} \textbf{2022}, \emph{16}, 5258--5273\relax
\mciteBstWouldAddEndPuncttrue
\mciteSetBstMidEndSepPunct{\mcitedefaultmidpunct}
{\mcitedefaultendpunct}{\mcitedefaultseppunct}\relax
\EndOfBibitem
\bibitem[Zhang \latin{et~al.}(2018)Zhang, Liu, Zhao, Yu, Reisner, and
  Dunbar]{small2018}
Zhang,~Y.; Liu,~X.; Zhao,~Y.; Yu,~J.-K.; Reisner,~W.; Dunbar,~W.~B. Single
  Molecule DNA Resensing Using a Two-Pore Device. \emph{Small} \textbf{2018},
  \emph{14}, 1801890\relax
\mciteBstWouldAddEndPuncttrue
\mciteSetBstMidEndSepPunct{\mcitedefaultmidpunct}
{\mcitedefaultendpunct}{\mcitedefaultseppunct}\relax
\EndOfBibitem
\bibitem[Pud \latin{et~al.}(2016)Pud, Chao, Belkin, Verschueren, Huijben,
  Van~Engelenburg, Dekker, and Aksimentiev]{pud2016mechanical}
Pud,~S.; Chao,~S.-H.; Belkin,~M.; Verschueren,~D.; Huijben,~T.;
  Van~Engelenburg,~C.; Dekker,~C.; Aksimentiev,~A. Mechanical trapping of DNA
  in a double-nanopore system. \emph{Nano letters} \textbf{2016}, \emph{16},
  8021--8028\relax
\mciteBstWouldAddEndPuncttrue
\mciteSetBstMidEndSepPunct{\mcitedefaultmidpunct}
{\mcitedefaultendpunct}{\mcitedefaultseppunct}\relax
\EndOfBibitem
\bibitem[Lubensky and Nelson(1999)Lubensky, and Nelson]{lubensky1999driven}
Lubensky,~D.~K.; Nelson,~D.~R. Driven polymer translocation through a narrow
  pore. \emph{Biophysical journal} \textbf{1999}, \emph{77}, 1824--1838\relax
\mciteBstWouldAddEndPuncttrue
\mciteSetBstMidEndSepPunct{\mcitedefaultmidpunct}
{\mcitedefaultendpunct}{\mcitedefaultseppunct}\relax
\EndOfBibitem
\bibitem[Bishop and Nasrabadi(2006)Bishop, and Nasrabadi]{bishop2006pattern}
Bishop,~C.~M.; Nasrabadi,~N.~M. \emph{Pattern recognition and machine
  learning}; Springer, 2006; Vol.~4\relax
\mciteBstWouldAddEndPuncttrue
\mciteSetBstMidEndSepPunct{\mcitedefaultmidpunct}
{\mcitedefaultendpunct}{\mcitedefaultseppunct}\relax
\EndOfBibitem
\bibitem[Hecht(2012)]{MR3043640}
Hecht,~F. New development in FreeFem++. \emph{J. Numer. Math.} \textbf{2012},
  \emph{20}, 251--265\relax
\mciteBstWouldAddEndPuncttrue
\mciteSetBstMidEndSepPunct{\mcitedefaultmidpunct}
{\mcitedefaultendpunct}{\mcitedefaultseppunct}\relax
\EndOfBibitem
\bibitem[Smith \latin{et~al.}(1996)Smith, Perkins, and Chu]{choudiff}
Smith,~D.~E.; Perkins,~T.~T.; Chu,~S. Dynamical Scaling of DNA Diffusion
  Coefficients. \emph{Macromolecules} \textbf{1996}, \emph{29},
  1372--1373\relax
\mciteBstWouldAddEndPuncttrue
\mciteSetBstMidEndSepPunct{\mcitedefaultmidpunct}
{\mcitedefaultendpunct}{\mcitedefaultseppunct}\relax
\EndOfBibitem
\bibitem[Ling and Ling(2013)Ling, and Ling]{ling2013distribution}
Ling,~D.~Y.; Ling,~X.~S. On the distribution of DNA translocation times in
  solid-state nanopores: an analysis using Schr{\"o}dinger’s
  first-passage-time theory. \emph{Journal of Physics: Condensed Matter}
  \textbf{2013}, \emph{25}, 375102\relax
\mciteBstWouldAddEndPuncttrue
\mciteSetBstMidEndSepPunct{\mcitedefaultmidpunct}
{\mcitedefaultendpunct}{\mcitedefaultseppunct}\relax
\EndOfBibitem
\bibitem[Carson \latin{et~al.}(2014)Carson, Wilson, Aksimentiev, and
  Wanunu]{carson2014}
Carson,~S.; Wilson,~J.; Aksimentiev,~A.; Wanunu,~M. Smooth DNA Transport
  through a Narrowed Pore Geometry. \emph{Biophysical Journal} \textbf{2014},
  \emph{107}, 2381--2393\relax
\mciteBstWouldAddEndPuncttrue
\mciteSetBstMidEndSepPunct{\mcitedefaultmidpunct}
{\mcitedefaultendpunct}{\mcitedefaultseppunct}\relax
\EndOfBibitem
\bibitem[Sarabadani and Ala-Nissila(2018)Sarabadani, and
  Ala-Nissila]{Sarabadani_2018}
Sarabadani,~J.; Ala-Nissila,~T. Theory of pore-driven and end-pulled polymer
  translocation dynamics through a nanopore: an overview. \emph{Journal of
  Physics: Condensed Matter} \textbf{2018}, \emph{30}, 274002\relax
\mciteBstWouldAddEndPuncttrue
\mciteSetBstMidEndSepPunct{\mcitedefaultmidpunct}
{\mcitedefaultendpunct}{\mcitedefaultseppunct}\relax
\EndOfBibitem
\bibitem[Kowalczyk \latin{et~al.}(2012)Kowalczyk, Wells, Aksimentiev, and
  Dekker]{kowalczyk2012slowing}
Kowalczyk,~S.~W.; Wells,~D.~B.; Aksimentiev,~A.; Dekker,~C. Slowing down DNA
  translocation through a nanopore in lithium chloride. \emph{Nano letters}
  \textbf{2012}, \emph{12}, 1038--1044\relax
\mciteBstWouldAddEndPuncttrue
\mciteSetBstMidEndSepPunct{\mcitedefaultmidpunct}
{\mcitedefaultendpunct}{\mcitedefaultseppunct}\relax
\EndOfBibitem
\bibitem[Reisner \latin{et~al.}(2009)Reisner, Larsen, Flyvbjerg, Tegenfeldt,
  and Kristensen]{reisnerpit}
Reisner,~W.; Larsen,~N.~B.; Flyvbjerg,~H.; Tegenfeldt,~J.~O.; Kristensen,~A.
  Directed self-organization of single DNA molecules in a nanoslit via embedded
  nanopit arrays. \emph{Proceedings of the National Academy of Sciences}
  \textbf{2009}, \emph{106}, 79--84\relax
\mciteBstWouldAddEndPuncttrue
\mciteSetBstMidEndSepPunct{\mcitedefaultmidpunct}
{\mcitedefaultendpunct}{\mcitedefaultseppunct}\relax
\EndOfBibitem
\bibitem[Klotz \latin{et~al.}(2015)Klotz, Duong, Mamaev, de~Haan, Chen, and
  Reisner]{klotzpit}
Klotz,~A.~R.; Duong,~L.; Mamaev,~M.; de~Haan,~H.~W.; Chen,~J. Z.~Y.;
  Reisner,~W.~W. Measuring the Confinement Free Energy and Effective Width of
  Single Polymer Chains via Single-Molecule Tetris. \emph{Macromolecules}
  \textbf{2015}, \emph{48}, 5028--5033\relax
\mciteBstWouldAddEndPuncttrue
\mciteSetBstMidEndSepPunct{\mcitedefaultmidpunct}
{\mcitedefaultendpunct}{\mcitedefaultseppunct}\relax
\EndOfBibitem
\bibitem[Storm \latin{et~al.}(2005)Storm, Storm, Chen, Zandbergen, Joanny, and
  Dekker]{storm2005fast}
Storm,~A.~J.; Storm,~C.; Chen,~J.; Zandbergen,~H.; Joanny,~J.-F.; Dekker,~C.
  Fast DNA translocation through a solid-state nanopore. \emph{Nano letters}
  \textbf{2005}, \emph{5}, 1193--1197\relax
\mciteBstWouldAddEndPuncttrue
\mciteSetBstMidEndSepPunct{\mcitedefaultmidpunct}
{\mcitedefaultendpunct}{\mcitedefaultseppunct}\relax
\EndOfBibitem
\bibitem[Lu \latin{et~al.}(2011)Lu, Albertorio, Hoogerheide, and
  Golovchenko]{lu2011origins}
Lu,~B.; Albertorio,~F.; Hoogerheide,~D.~P.; Golovchenko,~J.~A. Origins and
  consequences of velocity fluctuations during DNA passage through a nanopore.
  \emph{Biophysical journal} \textbf{2011}, \emph{101}, 70--79\relax
\mciteBstWouldAddEndPuncttrue
\mciteSetBstMidEndSepPunct{\mcitedefaultmidpunct}
{\mcitedefaultendpunct}{\mcitedefaultseppunct}\relax
\EndOfBibitem
\bibitem[Plesa \latin{et~al.}(2015)Plesa, Van~Loo, Ketterer, Dietz, and
  Dekker]{plesa2015velocity}
Plesa,~C.; Van~Loo,~N.; Ketterer,~P.; Dietz,~H.; Dekker,~C. Velocity of DNA
  during translocation through a solid-state nanopore. \emph{Nano letters}
  \textbf{2015}, \emph{15}, 732--737\relax
\mciteBstWouldAddEndPuncttrue
\mciteSetBstMidEndSepPunct{\mcitedefaultmidpunct}
{\mcitedefaultendpunct}{\mcitedefaultseppunct}\relax
\EndOfBibitem
\bibitem[Hernandez-Garcia \latin{et~al.}(2012)Hernandez-Garcia, Werten, Stuart,
  de~Wolf, and De~Vries]{hernandez2012coating}
Hernandez-Garcia,~A.; Werten,~M.~W.; Stuart,~M.~C.; de~Wolf,~F.~A.;
  De~Vries,~R. Coating of single DNA molecules by genetically engineered
  protein diblock copolymers. \emph{Small} \textbf{2012}, \emph{8},
  3491--3501\relax
\mciteBstWouldAddEndPuncttrue
\mciteSetBstMidEndSepPunct{\mcitedefaultmidpunct}
{\mcitedefaultendpunct}{\mcitedefaultseppunct}\relax
\EndOfBibitem
\bibitem[Laohakunakorn \latin{et~al.}(2013)Laohakunakorn, Ghosal, Otto,
  Misiunas, and Keyser]{laohakunakorn2013dna}
Laohakunakorn,~N.; Ghosal,~S.; Otto,~O.; Misiunas,~K.; Keyser,~U.~F. DNA
  interactions in crowded nanopores. \emph{Nano letters} \textbf{2013},
  \emph{13}, 2798--2802\relax
\mciteBstWouldAddEndPuncttrue
\mciteSetBstMidEndSepPunct{\mcitedefaultmidpunct}
{\mcitedefaultendpunct}{\mcitedefaultseppunct}\relax
\EndOfBibitem
\end{mcitethebibliography}
\end{document}


\maketitle

\newpage
\section{Dwell time with different absolute voltage settings}

\begin{figure}[h]
    \centering
    \includegraphics[width=\linewidth]{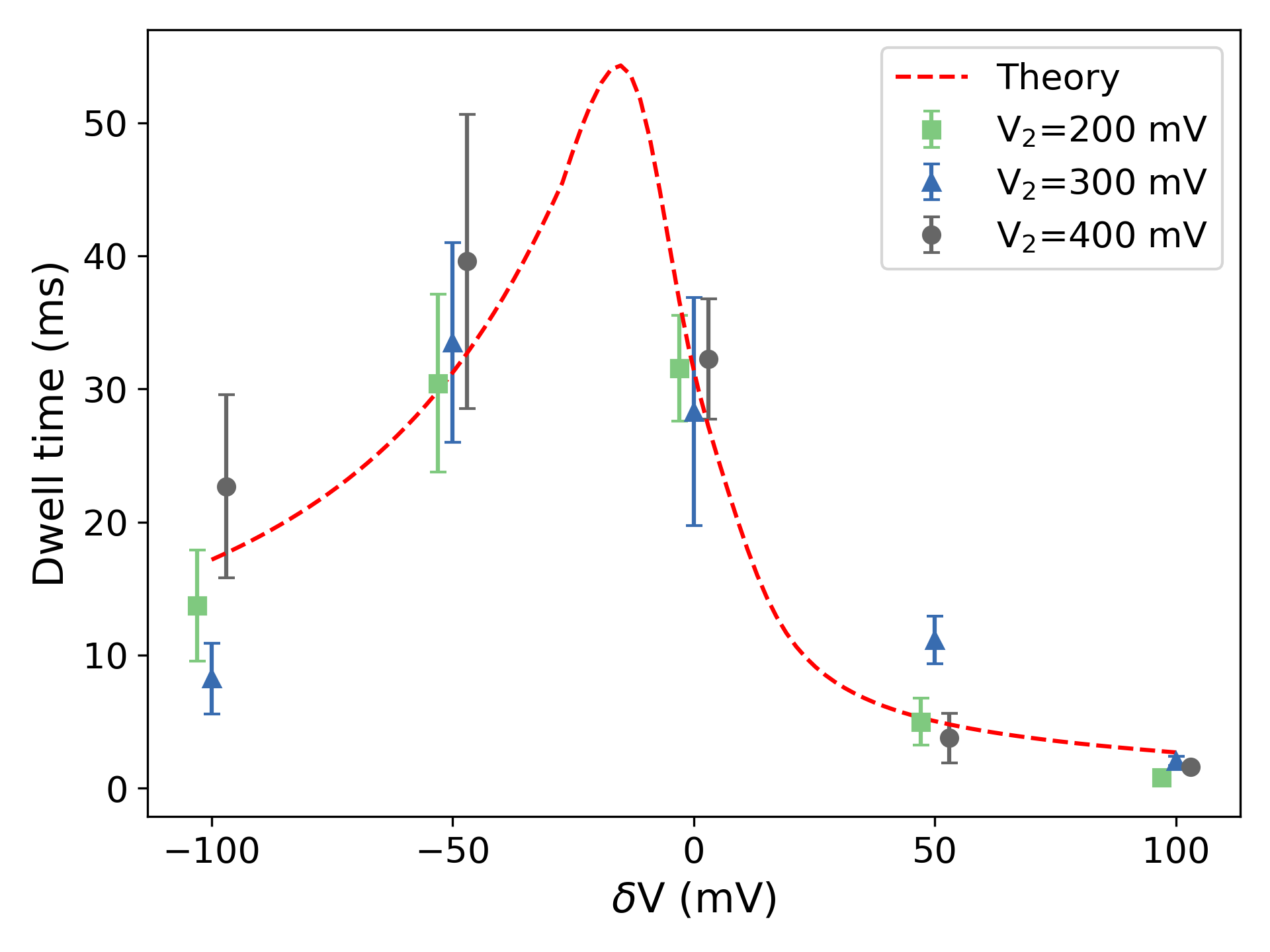}
    \caption{The dwell time of the T$_4$-DNA TOW with different absolute voltages. The red dashed line gives the theoretical curve using the fitted diffusion coefficient and electrophoretic mobility. The error bars give the standard error of the mean. }
    \label{fig:dwell_absv}
\end{figure}
We set the $V_2$ to 200\,mV, 300\,mV and 400\,mV and vary the differential voltage. The average dwell time from different absolute voltages fall within the standard error of the mean, appearing independent of the absolute voltage.

\newpage
\section{$\lambda$-DNA dwell time distribution.}
\begin{figure}
    \centering\includegraphics[width=\linewidth]{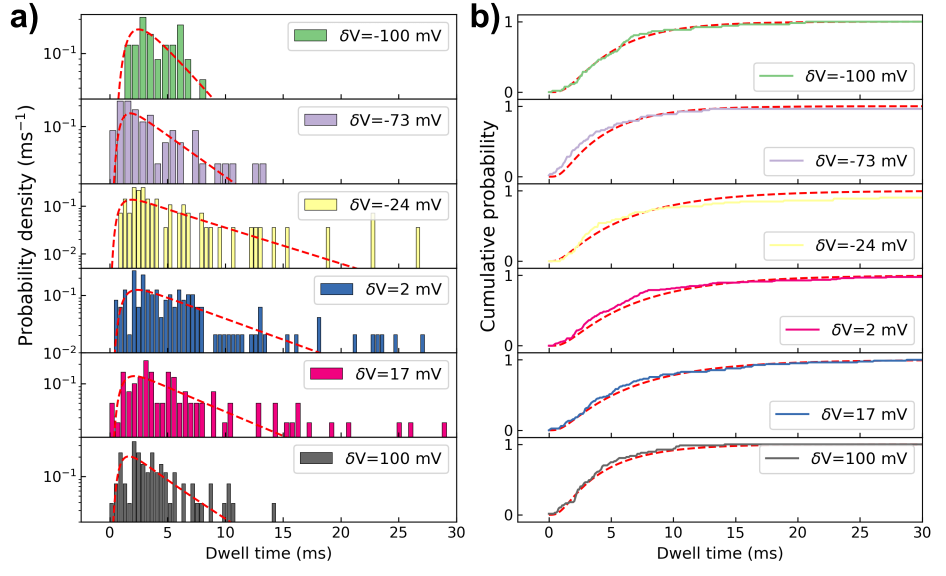}
    \caption{\textbf{a)} Histogrammed $\lambda$-DNA dwell times and \textbf{b)} corresponding cumulative distributions. The red dashed line indicates model fits.}
    \label{fig:cumulative_dwell}
\end{figure}
We find that the $\lambda$-DNA dwell time distributions depend symmetrically on the polarity of $\delta$V. This dwell time distribution symmetry can be confirmed by the cumulative probability in Fig.\,\ref{fig:cumulative_dwell}.b).  We use 1000 bins for all $\delta V$ to remove any artifacts introduced by binning differences.

\newpage
\section{First-passage model for balanced and unbalanced TOW initial position}

\begin{figure}
    \centering
    \includegraphics[width=\linewidth]{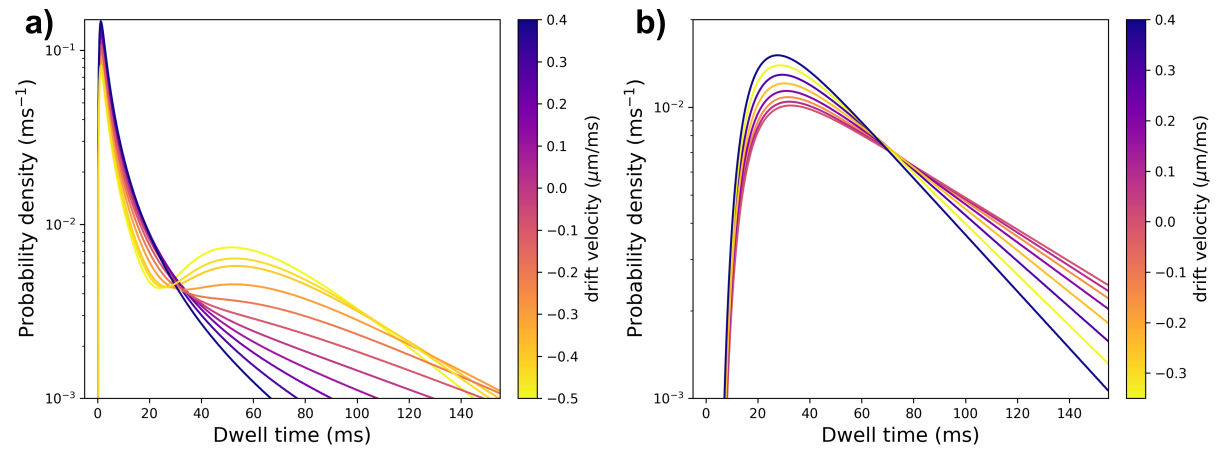}
    \caption{\textbf{a)} Unbalanced TOW initial position ($\alpha=0.9$) plotted for various values of drift velocity. \textbf{b)} Balanced TOW initial position ($\alpha=0.5$) plotted for various values of drift velocity.}
    \label{fig:model}
\end{figure}
We show the model dwell time distribution with unbalanced ($\alpha=0.9$) and balanced ($\alpha=0.5$) TOW initial position. For the unbalanced scenario, we observe the drift peak emerges as the drift velocity grows in the direction of P1 to P2 (with most DNA partitioned in the reservoir adjoining P1). For the balanced scenario, the dwell time distribution does not show a two-peak structure as we increase the drift velocity. The dwell time distribution is also symmetric as we change the polarity of $\delta V$. 
\newpage
\section{DNA free end velocity in presence of a second DNA strand}
\begin{figure}
    \centering
    \includegraphics[width=\linewidth]{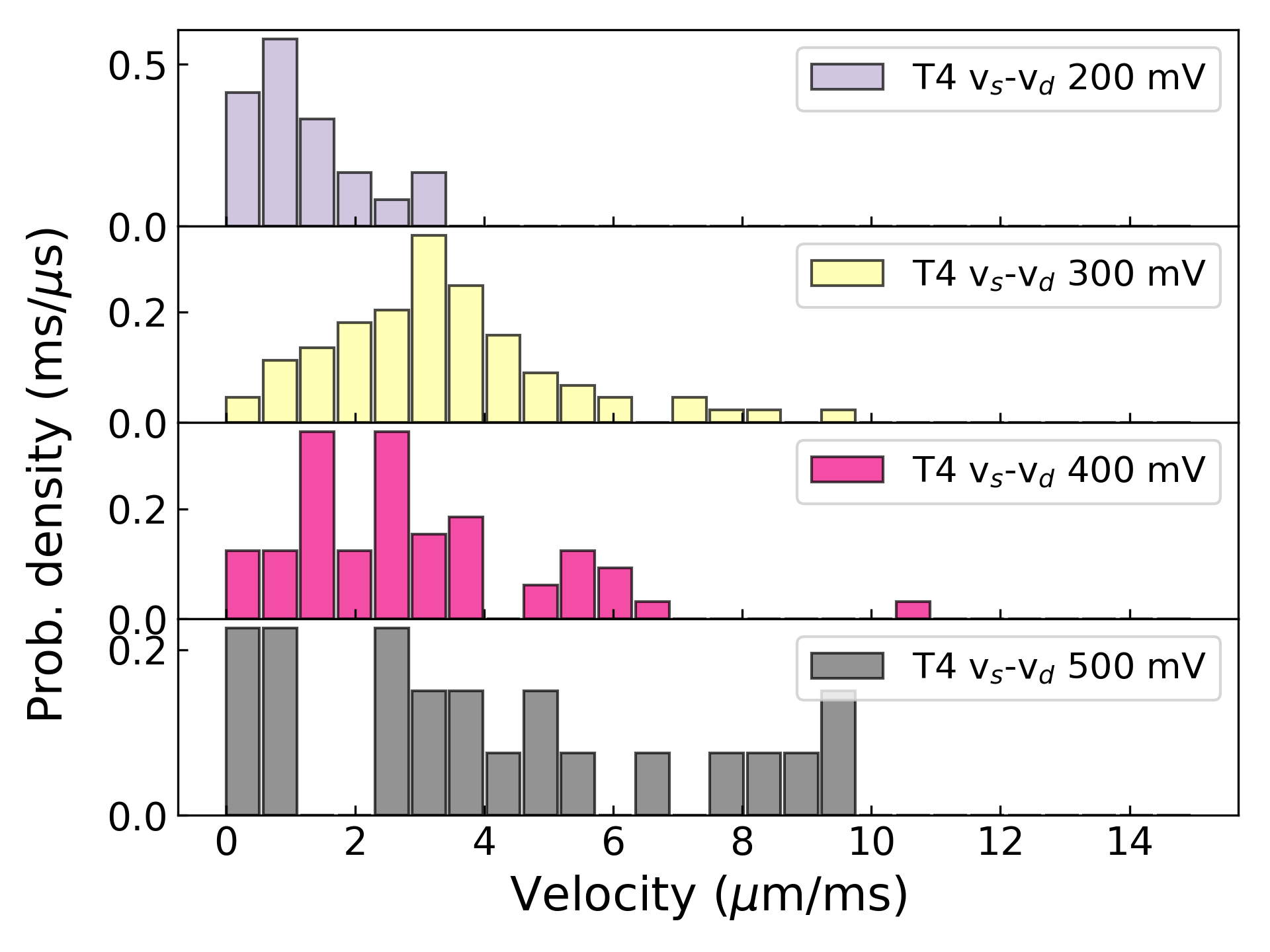}
    \caption{The velocity difference between $v_\mathrm{s}$ and $v_\mathrm{d}$ for different exiting pore voltages.}
    \label{fig:vsd}
\end{figure}
We observe with the same TOW event, $v_\mathrm{s}$ is \textit{always} greater than $v_\mathrm{d}$, indicating an additional friction caused by the presence of a second DNA strand from the fold.
\newpage

\section{Event counting for complex folding}

\begin{figure}
    \centering
    \includegraphics[width=\linewidth]{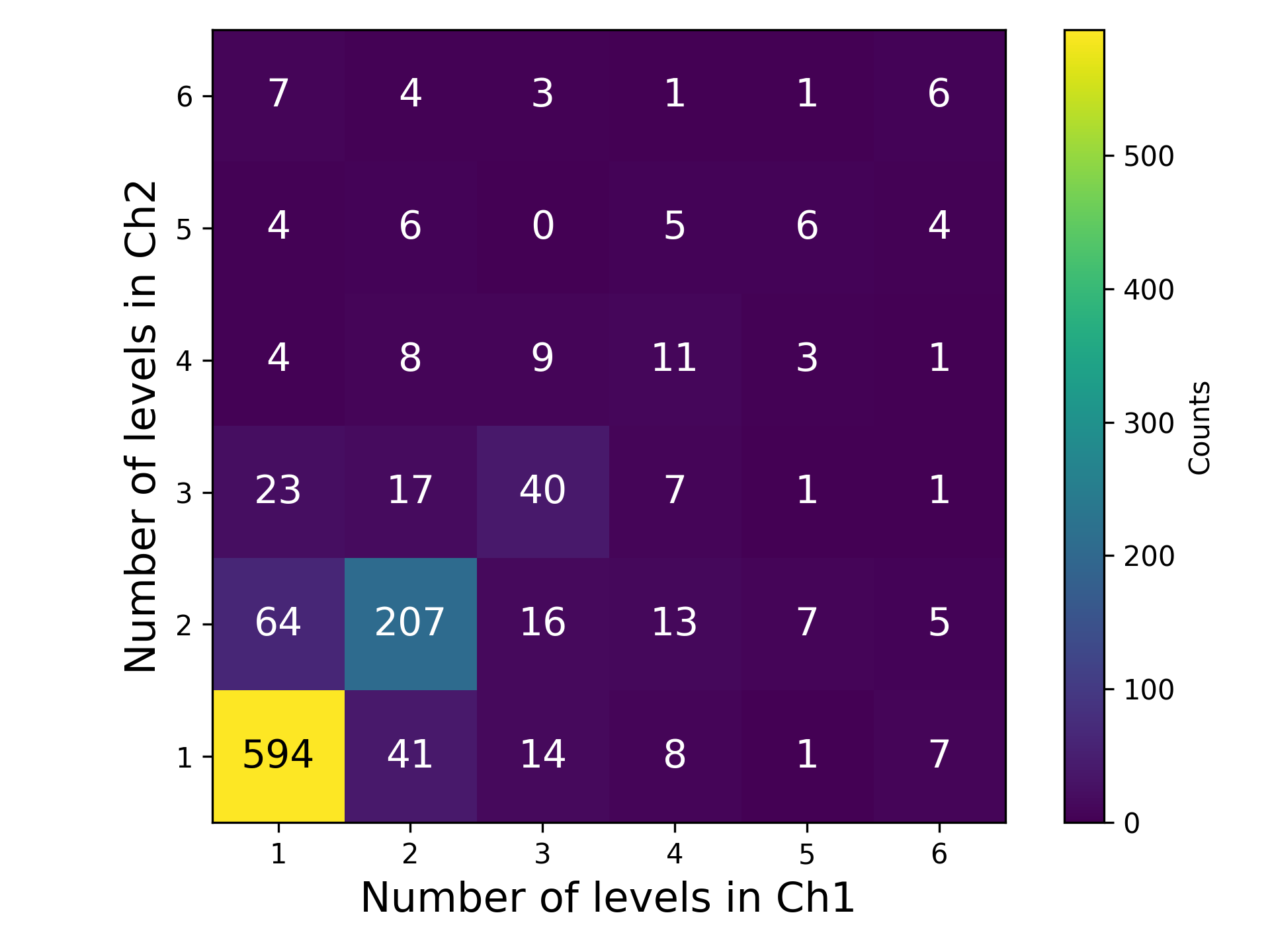}
    \caption{Number of events for T4-DNA with different degrees of folding as determined by the number of distinct levels in pore 1 channel (Ch1) and pore 2 channel (Ch2).  Data shown includes all voltage settings investigated.}
    \label{fig:complexfoldcounts}
\end{figure}

This section presents statistics of complex folding events observed in dual pore translocation.  The number of levels in Fig.~\ref{fig:complexfoldcounts} represent the maximum number of DNA strands present in the corresponding pore channel. For example, events with one level in both channels correspond to unfolded events. Events with one level in Ch1 and two levels in Ch2 correspond to half-folded molecules, in which the DNA free end resides in the common chamber. Events with 2 levels in each channel represent a single fold that spans the pores.   In general, folded events where the DNA free-end does not lie inside the common chamber are located along the matrix diagonal. The amounts of data for complex folding (more than a single fold) is statistically insufficient as there are less than 10 events for each voltage setting (for example, there are only 40 events  with three strands in each channel, corresponding to events with two folds).